\documentclass{aa} 

\usepackage{graphicx}
\graphicspath{{./}} 
\usepackage{epstopdf}
\usepackage[varg]{txfonts}

\begin{document} 

\title{Reconstruction of the ground-layer adaptive-optics point spread function for MUSE Wide Field Mode observations}
\subtitle{}
\titlerunning{GLAO PSF reconstruction for MUSE WFM}
\authorrunning{T. Fusco et al.}

\author{Thierry Fusco\inst{1,2}, Roland Bacon\inst{3}, Sebastian Kamann\inst{4}, Simon Conseil\inst{5,3}, Benoit Neichel\inst{2},
Carlos Correia\inst{2},
Olivier Beltramo-Martin\inst{1,2},
Joel Vernet\inst{6},
Johann Kolb\inst{6}
and Pierre-Yves Madec\inst{6}}

\institute{DOTA, ONERA, Université Paris Saclay, F-91123 Palaiseau, France \and 
 Aix Marseille Univ, CNRS, LAM, Laboratoire d'Astrophysique de Marseille, Marseille, France\ \and
 Univ. Lyon, Univ. Lyon1, ENS de Lyon, CNRS, Centre de Recherche Astrophysique de Lyon UMR5574, 69230 Saint-Genis-Laval, France\
 \and 
Astrophysics Research Institute, Liverpool John Moores University, 146 Brownlow Hill, Liverpool L3 5RF, United Kingdom \and 
Gemini Observatory/NSF's OIR Lab, Casilla 603, La Serena, Chile \and 
European Southern Observatory,Karl-Schwarzschild-Str. 2, 85748 Garching bei Muenchen,  Germany}

   \date{Received January 28, 2020; accepted February 25, 2020}
\abstract{Here we describe a simple, efficient, and most importantly fully operational point-spread-function(PSF)-reconstruction approach for laser-assisted ground layer adaptive optics (GLAO) in the frame of the Multi Unit Spectroscopic Explorer (MUSE) Wide Field Mode.}
{Based on clear astrophysical requirements derived by the MUSE team and using the  functionality of the current ESO Adaptive Optics Facility we aim to develop an operational PSF-reconstruction (PSFR) algorithm  and test it both in simulations and using on-sky data.}
{The PSFR approach is based on a Fourier description of the GLAO correction to which the specific instrumental effects of MUSE Wide Field Mode (pixel size, internal aberrations, etc.) have been added. It was first thoroughly validated with full end-to-end simulations. Sensitivity to the main atmospheric and AO system parameters was analysed and the code was re-optimised to account for the sensitivity found. Finally, the optimised algorithm was tested and commissioned using more than one year of on-sky MUSE data.}
{We demonstrate with an on-sky data analysis that our algorithm meets all the requirements imposed by the MUSE scientists, namely an accuracy better than a few percent on the critical PSF parameters including full width at half maximum and global PSF shape through the kurtosis parameter of a Moffat function.}
{The PSFR algorithm is publicly available and is used routinely to assess the MUSE image quality for each observation. It can be included in any post-processing activity which requires knowledge of the PSF.} 


\maketitle
%

\section{Introduction}

Achieved image quality is usually the primary parameter of successful observations, especially those performed at ground-based telescopes where atmospheric turbulence produces highly changeable conditions. 
For many scientific applications, precise knowledge of the achieved image quality is an absolute prerequisite. An obvious example is the comparison with higher spatial resolution space observations, like those obtained with the  Hubble Space Telescope, which achieve a ten times higher  resolution than classical ground-based observations in median seeing conditions. Source optimal extraction, source deblending, and image deconvolution are other examples where accurate knowledge of the point spread function (PSF) is needed \citep{Beltramo-Martin-2019,Fetick-2019a}. 

In natural seeing observations, the PSF full width at half maximum (FWHM) is often used to quantify the achieved image quality. Most modern ground-based telescopes are equipped with a seeing monitor which provides a real-time estimate of the FWHM. This is very convenient to get a rough estimate of the PSF, but it is usually not accurate enough in a number of  science applications. Firstly, the PSF is obtained at zenith and at a given wavelength, and these parameters are usually different from the airmass and wavelength of the observation. Secondly, the measurement is taken with a small telescope and does not take into account the relative outer-scale size of the turbulence with respect to the size of the telescope, or the image quality of the telescope plus instrument system.

The easiest and best method to obtain a good estimate of the PSF is to take an \textit{a posteriori} measurement of a bright, unresolved, and isolated source on the final image or data cube. This is an advantage as it takes into account all the possible effects that can alter the PSF: the atmospheric turbulence but also the instrument finite resolution, the detector sampling, and even some inaccuracy of the data reduction chain. For natural seeing observations, several more or less elaborated models exist, such as for example the Moffat function or the multi-Gaussian function \citep{Bendinelli1987, Trujillo2001, Infante2019}.
However, in some cases there are no bright point sources in the field of view because it is too small and/or is located at high galactic latitude where Galactic stars are rare. This is for example the case of deep-field observations like the Hubble Ultra Deep Field (UDF, \citealt{Bacon2017}). 

The main challenges for extra-galactic observations are two-fold. Firstly, extra-galactic observations require long exposures over the course of different nights. Variation of the PSF over the whole observation campaign can potentially be significant: for instance, in its Wide Field Mode (WFM), the Multi Unit Spectroscopic Explorer (MUSE) PSF can change by more than 100\% over several nights. At the same time, cosmological fields are usually devoid of point sources that can be used to monitor this variability  \citep{damjanov-2011}. Variation of the PSF then becomes a major limitation in kinematic or morphological analyses of distant galaxies. As such, Bouché et al. \citep{bouche-2015}, using state-of-the-art morpho-dynamical 3D algorithms, show that the PSF FWHM must be known to better than 20\% so as not to degrade the velocity parameters (maximum velocity, dispersion) by more than 10\%. These latter authors also highlight that the shape of the PSF (ellipticity) is critical for morphological parameters such as the inclination. There is a known degeneracy between rotational velocity and inclination of the system \citep{wright-2009,epinat-2010}, and as such, the PSF ellipticity must be known to 10\% or better. In this case, analytical PSF models are sufficient and better knowledge of the PSF is not critical as the accuracy of the analysis is limited by the morpho-kinematical model and/or signal-to-noise ratio.

In some other cases where the density of stars is too high, like in the centres of Globular Clusters, there are no isolated stars and more sophisticated techniques are needed to infer the PSF \citep{Kamann-2017,Kamann-2018}.
Even when a bright and isolated star is present in the field of view, the method described above assumes a uniform PSF over the field of view which is not always the case. We note that for seeing-limited observations with a limited field of view, the atmospheric  and  telescope PSF can indeed be considered as uniform with the field of view, but this is not generally the case for the instrument. 
With the generalisation of advanced adaptive optics (AO) systems, modern ground-based telescopes now offer improved image quality with respect to the seeing characteristics of the telescope site. However, the AO PSF is no longer a simple function of a single atmospheric parameter (seeing) but it is a complex function of the atmospheric turbulence (e.g. the profile of the atmospheric turbulence, the coherence time, and the anisoplanetic angle), the number of actuators, the accuracy and speed of the deformable mirror and the wave-front measurements, the brightness and location of the natural and laser guide stars, and the wavelength of observations. 

Despite this apparent complexity, AO systems offer a unique advantage in that they measure the atmospheric turbulence in real time at the exact location of the observation and through the same system used for the scientific observation. Thanks to the wave-front sensing telemetry information and good knowledge of the system, it is theoretically possible to predict the PSF, even without a point source within the field of view \citep{Veran-1997}.

Although PSF-reconstruction algorithms have been in existence for a long time \citep{Veran-1997,Gendron-2006,Gilles-2012,Ragland-2016,Beltramo-Martin-2019}, most of them are too complex to implement and  are not robust enough to be used blindly in normal operations. The fact that AO techniques have also evolved rapidly in parallel giving birth to a large number of species (e.g. single conjugate AO, ground layer AO, laser tomography AO, multi conjugate AO) has also prevented the development of robust and stable PSF-reconstruction algorithms and their validation on sky.
However, today the situation has changed  with the advent of the ESO Adaptive Optics Facility (AOF,\citealt{Arsenault-2008,Oberti2018}) at the Very Large Telescope (VLT) which is now in regular operation at UT4 since 2017 with the HAWK-I \citep{Pirard2004} and MUSE \citep{Bacon2004, Bacon2014} instruments.

With the regular use of MUSE Ground Layer AO (GLAO) mode, the number of non-AO expert users has increased significantly, and the need for an efficient and robust  PSF-reconstruction system is becoming more and more important. Another motivation is the need to qualify and rank the observations during service mode operation. The previous scheme based on the seeing monitor information cannot be used, as other critical information such as ground-layer fraction must be taken into account.

Here, we present a PSF-reconstruction algorithm developed specifically for the GLAO mode of MUSE. The paper is organised as follows. After a brief presentation of the MUSE instrument and its wide-field mode, we describe the AOF module that allows users to correct for the ground-layer contribution of the atmosphere, significantly improving the MUSE final images. We then present our PSF reconstruction scheme, its specificity, and its optimisation with respect to the typical performance of MUSE-WFM and AOF-GLAO correction. A sensitivity analysis using End-to-End simulations is provided and some algorithm parameters are then adjusted accordingly. After demonstrating the algorithm performance on simulated (and thus well-mastered) data, it is applied to a real MUSE on-sky observation. Thanks to more than one year of Globular Cluster data, we are able to provide a statistical analysis of our algorithm in operational conditions. Final results of this study are reported here with a clear demonstration of the efficiency of our final PSF-reconstruction (PSFR) approach. The final implementation in the MUSE pipeline is detailed and a first astrophysical application to the MUSE Ultra Deep Field observation is presented as an illustration of the importance and the power of the unique combination of PSF reconstruction and GLAO corrected wide field MUSE 3D cubes \citep{Bacon2017}. 

\section{MUSE Wide Field Mode}
MUSE is the ESO VLT second-generation wide-field integral field spectrograph operating in the visible \citep{Bacon-2014},  covering a simultaneous spectral range of 480-930 nm with a spectral resolution of $\sim$3000. Its Wide Field Mode (WFM) offers a field of view of 1 arcmin$^2$, sampled at 0.2\farcs. MUSE is composed of 24 identical channels, each one comprising a single Integral Field Unit (IFU) with an image slicer, a spectrograph, and a 4k$\times$4k CCD. MUSE has been in regular operation since October 2014. It was used in natural seeing mode until October 2017 when its GLAO mode was als made available to the ESO community. 

\subsection{GALACSI-GLAO and its specificities}\label{sec:GLAOerrorbud}
The MUSE GLAO mode is performed by the Ground Atmospheric Layer Adaptive Optics for Spectroscopic Imaging (GALACSI) and is part of the AOF, a full AO system with a deformable secondary mirror of 1170  actuators, four 20-Watt laser guide stars, and two wave-front sensing units (GALACSI and GRAAL) at each Nasmyth Platform, feeding MUSE and Hawk-I, respectively. Commissioned in 2017, GALACSI provides improved image quality (e.g. 10-50\% improved FWHM) in the MUSE wavelength range (480 - 930 nm) and over the full 1 arcmin$^2$ field of view of MUSE\citep{Kolb-2017}. The system is robust enough to now be the `normal' mode of operation of MUSE wide field mode. 
The MUSE WFM image-quality requirements, mainly driven by the instrument sampling and field of view (FoV), have led to a very specific correction stage which is focused on ground-layer correction. Hence, it does not try to reach the telescope diffraction limit but rather to improve the `equivalent' seeing \citep{Oberti2018,MUSE-WFM-spec}. In that respect, both the AO error budget breakdown and the PSF shape are very different from the classical AO ones. In addition, the performance criterion is no longer the Strehl Ratio but rather some parameters related to the PSF shape such as its FWHM or ensquared energy in various box sizes. Even though these parameters could be related to the residual wave-front variance, the relation is far less obvious than the for the Strehl Ratio and this new paradigm  in terms of PSF shape and performance has to be analysed and taken into account in the PSF reconstruction scheme. 

Analysis of the non-AO PSF on MUSE sky data has shown that a circular MOFFAT gives an accurate description of the PSF core and wing \citep{Moffat1969,Andersen2006,Muller2006}. The smooth evolution of the Moffat shape  parameters (FWHM, $\beta$) with wavelength can also be fitted with a low-order polynomial. This model has been extensively used with success for science analysis since MUSE began operation and its relevance has been fully demonstrated. From this basis and because the GLAO mode only provides a very partial correction and the resulting PSF is far from being diffraction limited. GLAO correction can be seen as a seeing improvement and in that respect a simple yet efficient way to describe a GLAO-corrected PSF is still to consider a Moffat function that can be fully described by its FWHM and its kurtosis ($\beta$ coefficient which characterises the wing shape of the PSF). The FWHM could be non-symmetrical if we need to account for residual anisoplanatism effects. More details and a justification of the Moffat choice for the GLAO PSF parameters is given in Sect. \ref{sec:psf-param}. 

\begin{figure}[ht!]
\includegraphics[width=1\linewidth]{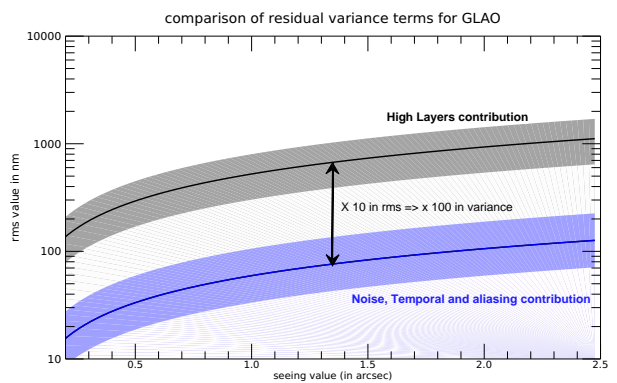}
\caption{Evolution of $\sigma^2_{\text{High Layers contrib}}$ and other AO error terms ($ \sigma^2_{\text{noise}}, \sigma^2_{\text{aliasing}} \text{\ and \ } \sigma^2_{\text{tempo}}$) as a function of seeing. The grey and blue areas correspond to the possible variations of atmospheric parameters (wind speed, $C_n^2$ profiles, outer scales) for the various error items.}
\label{fig:var-AO}
\end{figure}

For the sake of simplicity and clarity, let us now focus on the FWHM and derive an error budget for a typical MUSE WFM PSF. 
\begin{equation}
FWHM_{\text{Final}} = \sqrt{FWHM^2_{\text{Tel}} + FWHM^2_{\text{Atm, GLAO}} +FWHM^2_{\text{MUSE}}} \label{eq:full-error-budget}
\end{equation}
$FWHM_{\text{Tel}}$ stands for the telescope diffraction and any defect related to its aberrations that will not be corrected by the AO stage (high-temporal-frequency wind shake and/or vibrations, field aberrations, etc.); $FWHM_{\text{Atm, GLAO}}$ stands for the FWHM extension due to the uncorrected part of the atmospheric residual phase and $FWHM_{\text{MUSE}}$ stands for the FWHM extension due to the MUSE configuration: its coarse sampling and its own internal optical aberrations (not corrected by the AO loop). 
In the following we assume that $FWHM_{Tel}$ and $FWHM_{\text{MUSE}}$ are constant whatever the observation, whereas $FWHM_{\text{Atm, GLAO}}$ is a time-dependant contribution (this assumption is discussed in Sect. \ref{sec:MUSE-intern}. 
There is no simple, analytical way to link $FWHM_{\text{Atm, GLAO}}$ and $\sigma^2_{\text{Atm, GLAO}}$ (the residual variance after AO correction). Nevertheless, it is straightforward to say that they follow the same monotonic behaviour. In other words, being able to identify the dominant terms of the residual variance will give us critical items for developing an efficient model of the GLAO part of the MUSE PSF.  In that respect, a full GLAO error budget can be developed as follows:
\begin{equation}\label{eq:GLAO-error_budget}
\sigma^2_{\text{Atm, GLAO}}   = \sigma^2_{\text{High Order modes}} + \sigma^2_{\text{Tip tilt}} 
,\end{equation}
\begin{eqnarray}\label{eq:GLAO-HO}
\text{with \  }\sigma^2_{\text{High Order modes}} & =  & \sigma^2_{\text{fitting}} +\sigma^2_{\text{aliasing}} + \sigma^2_{\text{High Layers contrib}} \nonumber \\
                                   &  + & \sigma^2_{\text{noise}} + \sigma^2_{\text{tempo}} 
,\end{eqnarray}
\begin{eqnarray}\label{eq:GLAO-LO}
\text{and \  }\sigma^2_{\text{Tip tilt}} & =  & \sigma^2_{\text{TT, aliasing}} + \sigma^2_{\text{TT, anisoplanatism}} \nonumber \\
                                   &  + & \sigma^2_{\text{TT, noise}} + \sigma^2_{\text{TT, tempo}}
\end{eqnarray}
In the above error budget list, which gathers all the known error terms for such an instrument (due to spatial sampling, measurement noises, temporal error, anisoplanatism, etc.),  there are several points worth highlighting: The fitting error mainly acts on high spatial frequencies, that is, those higher than the AO cut-off frequency defined as $1/(2 d)$, with $d$ being the deformable mirror (DM) spacing on the PSF wings for example. 
The laser guide stars (LGS) are bright enough (20 Watts emitted on sky) to neglect the measurement noise on LGS WFS ($ \sigma^2_{\text{noise}} \simeq 0 $). 
As shown in Fig.~\ref{fig:var-AO}, $\sigma^2_{\text{High Layers contrib}}$ is typically 100 times larger than the other error terms (aliasing and temporal effects).
 
The tip tilt (TT) contribution can be decomposed into several terms among which noise and anisoplanatism are by far the dominant ones. The choice of natural guide star (NGS) in the technical FoV is mainly driven by its limit magnitude. The WFS characteristics could also be adapted to accommodate low flux NGS by changing the integration time. These two combined aspects lead to observing configurations where the noise term is never dominant in the error budget. The noise is neglected in the following; we note that it would have been straightforward to take it into account in our algorithm by adding a classical noise measurement and noise propagation term. A combination of simulations, AOF design, and AOF on-sky data shows that such an addition is of no real benefit in the MUSE WFM case. 

\begin{figure}[ht!]
\includegraphics[width=1\linewidth]{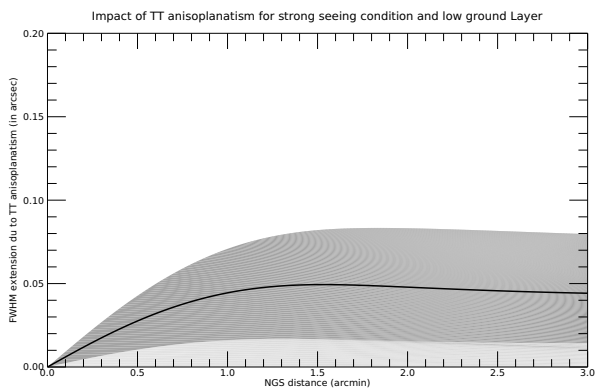}
\caption{Tip tilt anisoplanatism for various atmospheric conditions obtained using data gathered on the MUSE RTC after one year of observations. The solid line corresponds to an average profile, and the grey area corresponds to the scattering of more than 400 data points gathered during more than one year by the GALACSI RTC.}
\label{fig:TT-aniso}
\end{figure}

By design the NGS is always further than 1 arcmin from the optical axis. In that case, a good approximation assuming a two-layer model for the turbulence is to consider full decorrelation of the high-layer contribution and a full TT correction of the ground layer.  This is confirmed by Fig. \ref{fig:TT-aniso} where the TT anisoplanatism contribution is plotted for various atmospheric conditions. It is shown that the decorrelation hypothesis is very well validated but also that the final TT contribution due to the anisoplanatism effects remains very small (typically smaller than 50 mas) with respect to the MUSE pixel size (200 mas).

In all cases, the full decorrelation of TT anisoplanatism combined with the small contribution of TT noise leads to a non-elongated PSF (this has been experimentally confirmed on all the MUSE WFM images since the beginning of the instrument operation more than three years ago). The TT star is only here to ensure that ground-layer contributions of the atmosphere, the telescope pointing, and wobble aspects are corrected for.

The analysis of the various error terms clearly shows that in the specific case of GLAO, the PSF is mainly impacted by three  terms: the `fitting' and `high layer (HL) contribution' terms for the LGS (i.e. the High Order mode correction),  and the TT anisoplantism for the NGS contribution.
The latter three terms are the only ones considered in the following in our GLAO PSFR algorithm. 

\subsection{End-to-End simulation of the MUSE ground-layer adaptive-optics system}
End-to-End (E2E) simulations are carried out with  Object–Oriented Matlab Adaptive Optics - OOMAO  \citep{OOMAO}, which is a Matlab community-driven toolbox dedicated to AO systems. The OOMAO toolbox is based on a set of classes representing the source, atmosphere, telescope, wave-front sensor(WFS), Deformable Mirror (DM), and an imager of an AO system. It can simulate NGS and LGS single-conjugate AO (SCAO) and tomography AO systems on monolithic and/or segmented telescopes up to the size of the Extremely Large Telescope (ELT). 

\begin{figure}[ht!]
\includegraphics[width=1\linewidth]{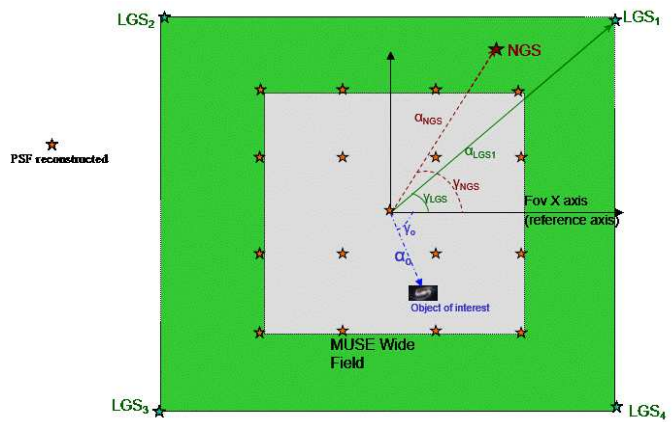}
\caption{MUSE WFM and AOF geometry in the FoV.}
\label{fig:config-MUSE}
\end{figure}

We used OOMAO for simulating the full AOF system in order to validate our PSFR algorithm and to provide a comprehensive sensitivity analysis of the PSFR performance. The simulation parameters (from the system and the environment view points) are listed below. Figure \ref{fig:config-MUSE} presents the problem geometry and more precisely the positions of LGS, NGS, and the directions of interest in the FoV. Tables \ref{tab:system_param} and \ref{tab:turb_param} present the system and atmospheric parameters used in the simulation and sensitivity analysis.

\begin{table*}[ht!]
\centering
\caption{System parameters used for the E2E simulation and the PSFR validation process.}
\label{tab:system_param}
\begin{tabular}{|l||l|l|}
\hline
Telescope & \multicolumn{2}{l|}{8m}\\\hline
Central Obstruction &  \multicolumn{2}{l|}{14\%}\\\hline
Deformable mirror  & \multicolumn{2}{l|}{DSM (deformable secondary mirror) 1100 actuators}\\\hline
\# of corrected modes & \multicolumn{2}{l|}{490}\\\hline
LGS WFS & LGS focus alt & 90 km \\ 
        & number of lenslets & 40x40\\
        & pixel size & 0.5" \\
        & frame rate & 1000Hz\\
        & wavelength & 589 nm \\
        & \# of pixels per sub-ap & 6x6\\ 
        & photons/sub-ap/frame  & 500 ph\\
        & ron   & 0.1e$^-$\\ \hline
TT WFS & full pupil imager & \\
        & number of pixels & 8x8\\ 
        & frame rate & 1000Hz\\
        & wavelength & H band \\
        & position & 120" off-axis \\
        & magnitude  & 15 \\
        & ron   & 10 e$^-$\\\hline
Loop parameters & loop frequency  & 1000Hz \\
                & GLAO computation & $\sum$ of LGS signal (TT-removed) \\
                & GLAO gain & 0.5 \\
\hline
\end{tabular}
\end{table*}

\begin{table*}[ht!]
\centering
\caption{Turbulence parameters used for the E2E simulation and the PSFR validation process.}
\label{tab:turb_param} 
\begin{tabular}{|c||c|c|c|c|c|c|c|c|c|c|}
\hline
seeing & \multicolumn{10}{l|}{0.8" @ 0.5$\mu m$} \\ \hline
outer scale & \multicolumn{10}{l|}{20m (same for each layer)} \\ \hline
$C^2_n(h)$ [in \%]  & 59 & 2 & 4 & 6 & 3 & 3 & 9 & 4 & 5 & 5 \\ \hline 
alt [in km]     & 0 & 0.1 & 0.3 & 0.6 & 1.1 & 2.3 & 4.5 & 7.8 & 11 & 14 \\ \hline
wind [in m/s]     & 6.6 & 5.9 & 5.1 & 4.5 & 5.1 & 8.3 & 16.3 & 30.2 & 34.3 & 17.5 \\ \hline
\end{tabular}
\end{table*}

\subsection{Choice of PSF model}\label{sec:psf-param}
One of the critical issues for any PSF reconstruction algorithm is the choice of PSF model. The most natural basis for describing the PSF is a pixel-wise basis. Although by definition, working on a pixel-wise basis removes the need for a model-based approximation, it is often not very well adapted to operational constraints because it requires a lot of memory and storage capacity. This is especially true for multi-wavelength instruments where one (or several if there are field variations to account for) PSF has to be computed and stored per wavelength bin. The choice of PSF model also strongly depends on the type of AO system (and thus AO correction) considered as well as the observation and post-processing requirements related to the astrophysical science cases. 
For MUSE WFM, the GLAO system only provides a very partial AO correction, and therefore the two critical parameters that have been identified to cover most of the science case requirements in terms of PSF knowledge are the PSF FWHM, which provides information on the data quality and final image resolution; and the level of the PSF wings, which provides information on the energy spread by the PSF in the FoV (spaxel contamination). 

\begin{figure}[ht!]
\includegraphics[width = 1\linewidth]{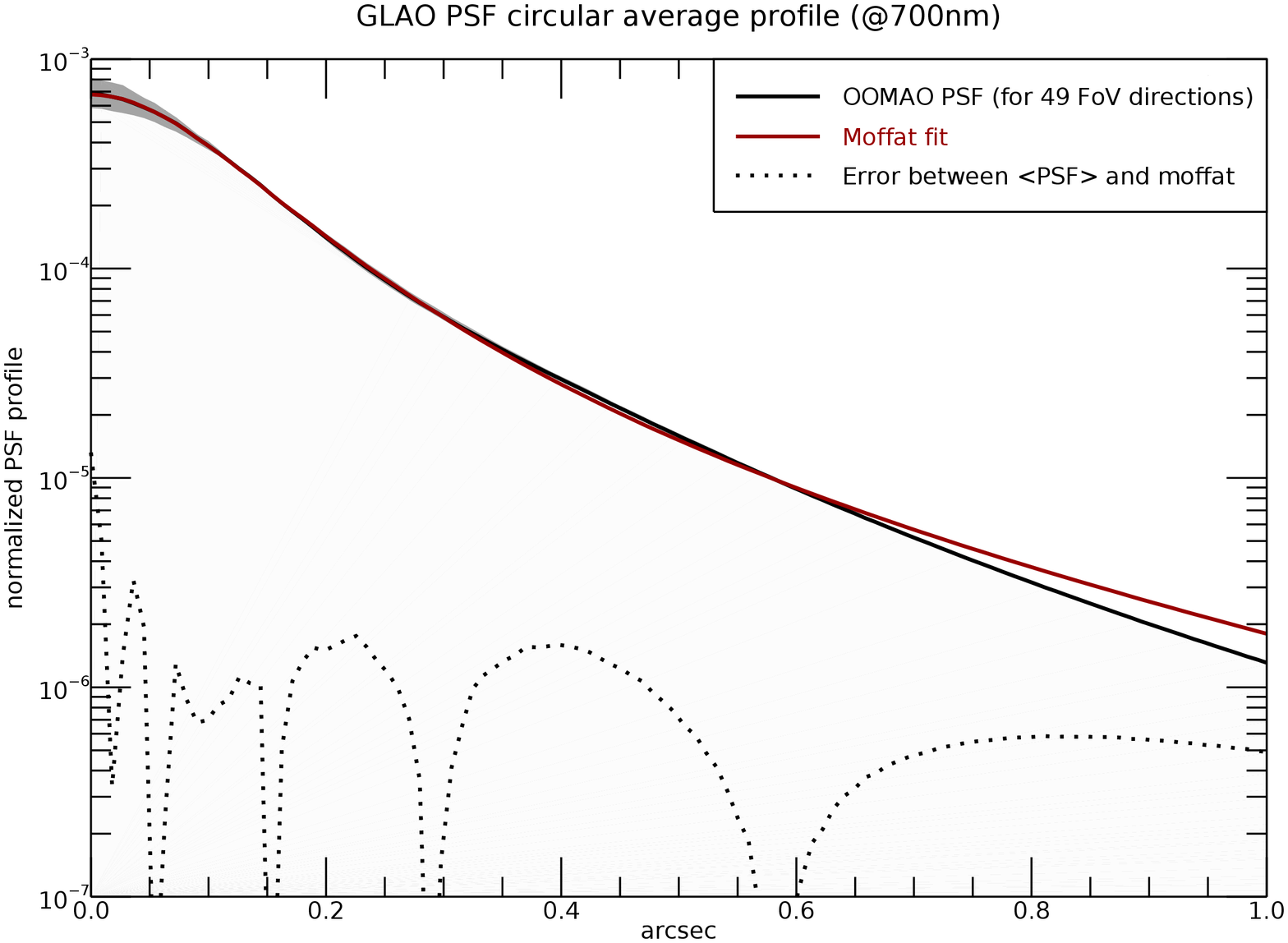}
\caption{Comparison of a GLAO PSF profile computed in the MUSE WFM. The black dots show 49 regularly spaced positions in 1x1 arcmin$^2$. The solid black line represents the average PSF for the 49 positions and the grey area shows the full dispersion of the PSF over the entire FoV. The red line shows the Moffat fit of the mean PSF.}
\label{fig:psf_comparison}
\end{figure}

Considering these two aspects and the typical shape of a very partially GLAO-corrected PSF, a Moffat model is particularly well adapted for the description of the MUSE WFM PSF. 
The Moffat PSF can be mathematically described as follows. 
\begin{equation}
 M(x,y) = M_0\left (\left ( \frac{x-m_x}{\alpha_x} \right )^2 + {\left ( \frac{y-m_y}{\alpha_y} \right )^2}+1\right )^{-\beta}
 ,\end{equation}
where $\alpha_x$ (resp. $y$) and $\beta$ are directly related to the function FWHM, $ M_0$ stands for the global amplitude factor, and $m_x,y$ for the absolute focal plane positions. Furthermore,
\begin{equation}
    FWHM_{x,y} = 2\alpha_{x,y}\sqrt{2^{\frac{1   }{\beta}}-1} 
,\end{equation}
where  $\beta$ is a very good marker of the shape of the PSF wings; the poorer the correction, the larger the $\beta$. It has been shown that a Moffat model with a $\beta$ value of greater than 4 is very well adapted for describing a purely turbulent PSF \citep{Trujillo2001}. Figure \ref{fig:psf_comparison} shows a comparison of a simulated GLAO PSF with OOMAO (and the nominal parameters defined above) with a Moffat fit of this PSF. 
We define a criterion on the PSF profile with respect to a given reference for a given focal plane area (s) as follows: 
\begin{equation}
Err_{ref,psf, s} = \sqrt{\frac{\iint_{-s/2}^{s/2}\left | PSF(x,y)-REF(x,y) \right |^2dxdy}{\iint_{-s/2}^{s/2}\left | REF(x,y)\right |^2dxdy}}*100 \label{eq:err}
,\end{equation}
where $REF(x,y)$ stands for the reference PSF (considered as the true one). This error parameter is used to evaluate the accuracy with which a Moffat can actually fit GLAO PSFs. Let us first focus on the Moffat description of the PSF. In that case, $Err_{<PSF>,Moffat,2*FWHM}$ (as defined in Eq \ref{eq:err}) is equal to 1.0, 1.1, and 2\% for imaging wavelengths of 500, 700, and 900 nm, respectively. This description of a GLAO PSF is therefore extremely accurate. The redder the wavelength, the better the correction, and therefore the more structured the PSF. This means that the model
errors will be greater for the larger wavelengths  than for smaller ones. Nevertheless an extensive analysis of the GLAO PSF in various atmospheric conditions and for the whole MUSE WFM spectral range shows that a Moffat fit always gives better results than a few percent which fully validates our choice of a Moffat description for the GLAO PSF.  
Let us now focus on the FoV evolution of the PSF. As mentioned above, we simulated 47 regularly spaced PSFs  in a 1x1 arcmin$^2$ FoV with our OOMAO simulator.  For each PSF, we fitted a Moffat function and we can therefore analyse the evolution of the Moffat parameters (FWHM and $\beta$).
\begin{table*}[ht!]
\caption{Evolution of PSF key parameters (FWHM and $\beta$) in the MUSE WFM FoV. Statistics were obtained using 47 regularly spaced PSFs in a 1x1 arcmin$^2$ FoV}
\label{tab:psf-parameters}
\centering
\begin{tabular}{|c||ccc|ccc|ccc|}
\hline
                & \multicolumn{3}{c|}{\bf 500 nm} & \multicolumn{3}{c|}{\bf 700 nm}  &\multicolumn{3}{c|}{\bf                                                                        900 nm} \\    
                & Average & RMS & PV   & Average & RMS & PV & Average & RMS & PV \\\hline
FWHM (arcsec)  &   0.318 & 0.016 & 0.05 &0.228 & 0.017 &  0.04 & 0.1844 & 0.017 & 0.05   \\\hline
$\beta$ &  1.66 & 0.1 &  0.3      &    1.52 & 0.1 &  0.2     &  1.55 & 0.1  & 0.3   \\\hline
\end{tabular}
\end{table*}
The FWHM rms error in the FoV is smaller than 20 mas and the value of $\beta$ is smaller than 0.1. From the previous simulation results we can consider a single PSF and apply it to the whole FoV. We note that, for further development, a more complex PSF model could be investigated. For example, R. Fetick \citep{Fetick-2019b} recently proposed a new PSF model for AO-corrected applications. This latter model relies on nine parameters and allows the user to fit an AO-corrected PSF both in its corrected area and its wings extremely accurately.

\section{Point-spread-function reconstruction for MUSE Wide Field Mode}
\begin{figure*}[ht!]
\centering
\includegraphics[width=1\linewidth]{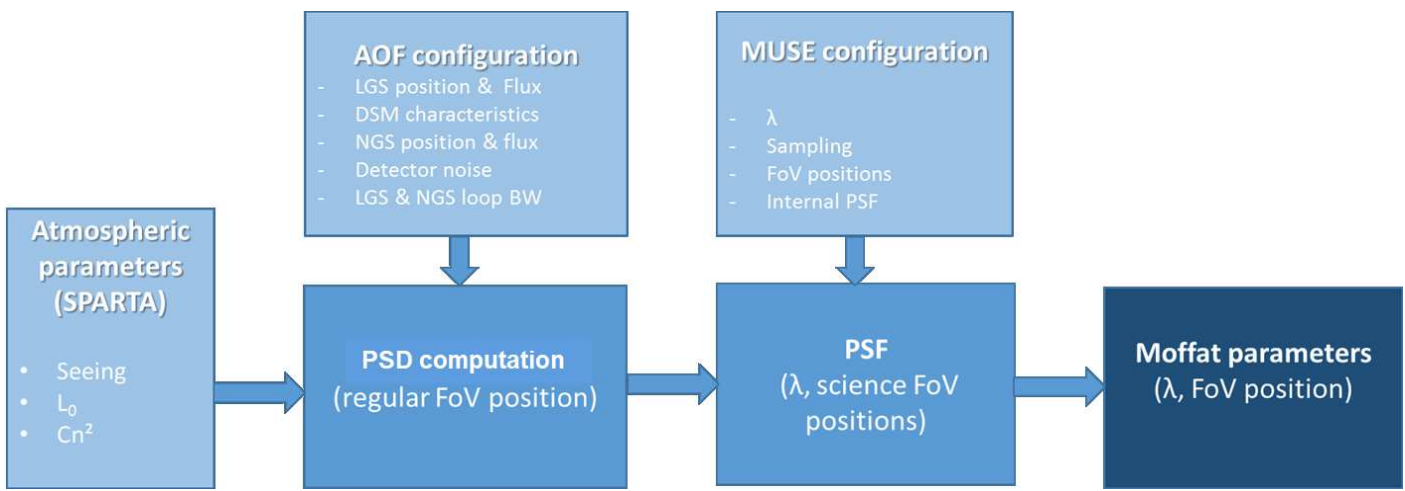}
\caption{Generic description of the PSFR algorithm.}
\label{fig:PSFR-algo-desc}
\end{figure*}

This section is dedicated to the presentation of the PSFR algorithm and its performance analysis on simulated data. Using the  output from the previous section, we now focus on the three Moffat parameters ($\alpha_x,\alpha_y$ and $\beta$) for each wavelength. One single PSF (resulting from the average of nine PSFs evenly distributed across the FoV) is estimated per wavelength bin. The whole PSFR process is summarised in Figure \ref{fig:PSFR-algo-desc}.
\subsection{The ground-layer adaptive-optics PSFR algorithm }
The starting point of the PSFR algorithm is to consider a Fourier  basis to describe the whole problematic. Here, it is assumed that everything (phase propagation, WFS measurements, DM commands) is linear and spatially shift-invariant. Hence, all the usual operators are diagonal with respect to spatial frequencies and simply act as spatial filters in the Fourier domain. It follows that each equation can be written frequency by frequency  \cite{Neichel2008}. The main advantage of the Fourier basis is its diagonal aspect  in the frequency domain. It follows that any  reconstruction algorithm may be derived and evaluated one Fourier component at a time. In addition, second-order statistics of the residual phase and long-exposure PSF can be evaluated directly without the need for iterations. By avoiding the convergence problem, simulation times are cut down by orders of magnitude at VLT scales.
\begin{equation}\label{eq:PSFR_nom}
\begin{aligned}
&PSF(x,y) = PSF_{Tel}\star PSF_{GLAO} \star PSF_{MUSE}\\
&PSF_{GLAO}(x,y) = FT^{-1} \left \{\exp{\left( -\frac{1}{2} FT\left \{ PSD_{\phi}(f_x,f_y,\lambda)\right \} \right)}\right \},
\end{aligned}
\end{equation}
where $PSD_{\phi}$ is the residual phase power spectral density after GLAO correction, $PSF_{Tel}$ is the telescope PSF defined by the telescope pupil, and $PSF_{MUSE}$ includes pixel effects and is defined as a centred Gaussian function with a FWHM of 0.2\farcs

The main limitation of the Fourier approach is that aperture-edge effects and boundary conditions that cannot be represented by shift-invariant spatial filters are neglected. Hence, the Fourier modelling only applies to the idealised case of an infinite aperture system, and all effects of incomplete beam overlap in the upper atmospheric layers are neglected. However, in the frame of MUSE WFM, the size of the telescope aperture is large enough (with respect to the sub-aperture diameter and to $r_0$) to satisfy this assumption. Moreover, the GLAO system and its simple averaging process is very well adapted to the Fourier representation: it allows the user to simply and directly focus on the dominant error terms in the error budget (fitting and high-altitude-layer contributions).

Using Equation \ref{eq:PSFR_nom},  PSFs are computed for each wavelength at nine positions in the FoV and then averaged. From the averaged PSF, a 2D Moffat fit is performed using a classical Least Square algorithm and the three main Moffat parameters $\alpha_x(\lambda_i)$,$\alpha_y(\lambda_i)$, and $\beta(\lambda_i)$ are stored for each $\lambda_i$ bin (for MUSE WFM, the number of bins is equal to 3000).  

\subsection{Sensitivity analysis}
A comprehensive analysis of the PSFR algorithm has been provided in close interaction with ESO and MUSE teams during the development process. Here we present a very small subset of the full analysis in order to illustrate the main conclusions. From the GLAO error budget presented in Section \ref{sec:GLAOerrorbud}, the dominant term from the performance point of view is the contribution of the uncorrected high-altitude layers ($\sigma^2_{\text{High Layers contrib}}$). This term depends on the three atmospheric parameters only:
$r_0 @ 0.5\mu m$ (or the seeing value $s = 0.1 / r_0 $ in arcsec); 
 ground layer fraction (GLF) - it is worth noting that the combination of GLF and  $r_0 @ 0.5\mu m$ gives the contribution of the uncorrected high turbulent layers - and $L_0$ (in m) which is the outer scale of the turbulence.

The combination of $r_0$ and 1-GLF (fraction of high-layer turbulence) gives the contribution of the uncorrected layers. In the following we study the impact of incorrect values for these three parameters on the PSFR performance. As proposed above, the system PSF is described by a Moffat function and we study the impact of the incorrect parameter values on both the Moffat FWHM and $\beta$. 

\begin{figure}[ht!]
\includegraphics[width=1\linewidth]{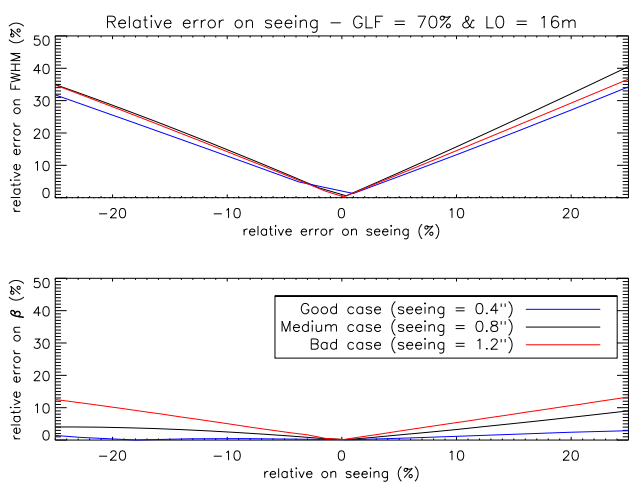}
\caption{Relative error on FWHM and $\beta$ (in \%)  as a function of the bias on seeing estimation. 3 cases of real seeing inputs are considered: Good case (0.4''), median case (0.8") and bad case (1.2"). the  GLF value is equal to 70\% (which roughly corresponds to the median values measured on 1 years of MUSE-WFM+GLAO operations) and $L_0$ is equal to 25m.}
\label{fig:psf_accuracy_simul_seeing_rel}
\end{figure}

Let us first focus on the seeing. Three cases are considered, assuming a $70\%$ GLF seeing and a 16m $L0$ (both extracted from typical/median values observed at Paranal): 
an optimistic case of 0.4" seeing condition;  a typical case of 0.8" seeing condition; and a pessimistic case of 1.2" seeing.
Figures  \ref{fig:psf_accuracy_simul_seeing_rel} shows the evolution of the error on FWHM and on $\beta$ as a function of an error on the  seeing estimation. 
Firstly, we can clearly see that the error on the FWHM estimation is strongly correlated with the error on the seeing with almost a one-to-one relationship. This can be easily understood in the case of partial correction (`typical' and `pessimistic' cases). In that case the high uncorrected turbulence is responsible for the broadening of the PSF and the PSF FWHM will be directly linked to the high altitude seeing which is given by the following relationship: 
\begin{equation}
seeing_{HL} = (1-GLF)^{3/5}*seeing
.\end{equation}
As shown previously, the GLAO correction does not only affect the FWHM of the PSF but also its shape (especially far from the optical axis). This shape is represented by the $\beta$ parameter in the case of a Moffat. Looking at  $\beta$ we can see that this parameter is significantly less affected by an error on the seeing parameter except for the optimistic case (when most of the turbulence is located near the ground). In that case, the correction becomes quite efficient and the shape of the PSF starts to change significantly and therefore the $\beta$ parameter starts to play a greater role in the overall PSF description.

Let us now consider an error on the GLF estimation. Here, again, three cases are considered (assuming a 0.8" seeing and a 16m $L_0$): an optimistic case where 90\% of the turbulence is located near the ground and is therefore corrected by the GLAO system. In that case the AO performance becomes quite important and a diffraction-limited core appears; 
a typical case where 70\% of the turbulence is located near the ground. In that case GLAO provides a significant reduction of the PSF FWHM without achieving the diffraction limit. This case is meant to represent the typical performance expected with the GLAO system for MUSE-WFM; 
a pessimistic case where a large amount (50\%) of turbulence remains uncorrected (in the high-altitude layers).

\begin{figure}[ht!]
\includegraphics[width=1\linewidth]{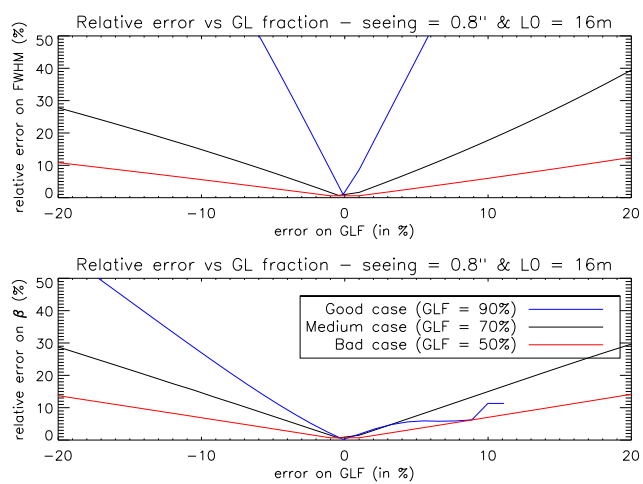}
\caption{Relative error on FWHM and $\beta$ (in \%)  as a function of the bias on GLF estimation. 3 cases of real GLF inputs are considered: Good case (90\%), median case (70\%) and bad case (50\%). the seeing value is equal to 0.8" (which roughly corresponds to the median values measured on 1 years of MUSE-WFM+GLAO operations) and $L_0$ is equal to 25m.}
\label{fig:psf_accuracy_simul_GL_rel}
\end{figure}
\begin{figure}[ht!]
\includegraphics[width=1\linewidth]{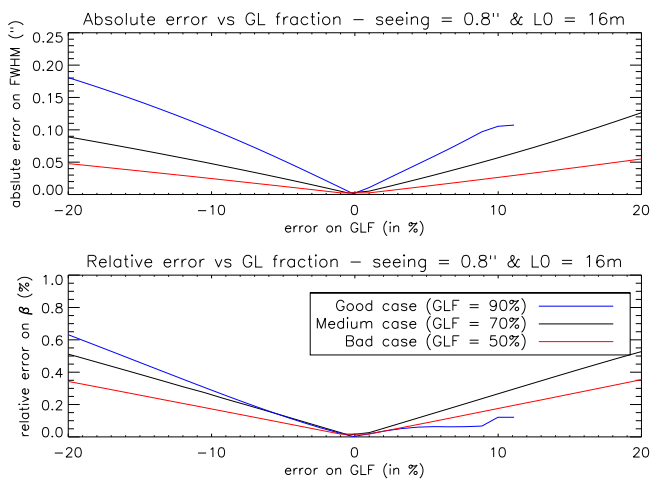}
\caption{Absolute error on FWHM (in arcsec) and $\beta$ (a.u.) as a function of the bias on GLF estimation. 3 cases of real GLF inputs are considered: Good case (90\%), median case (70\%) and bad case (50\%). the seeing value is equal to 0.8" (which roughly corresponds to the median values measured on 1 years of MUSE-WFM+GLAO operations) and $L_0$ is equal to 25m.}
\label{fig:psf_accuracy_simul_GL_abs}
\end{figure}

Figure \ref{fig:psf_accuracy_simul_GL_rel} shows the impact of an estimation error of the GLF (and thus of the high-layer uncorrected fraction of the turbulence) on the PSFR accuracy (looking at the Moffat parameters). Here again a linear behaviour between the error on the GLF and the estimated FWHM is found. It is worth noting that even though the relative error on FWHM and $\beta$ is relatively high for a small estimation error in the high GLF fraction case, the absolute values remain reasonable (see Figure \ref{fig:psf_accuracy_simul_GL_abs}). 

Errors on FWHM smaller than 50 mas are found for a GLF mis-estimation of typically $\pm 10 \%$. The "good case" (90 $\%$ of the turbulence near the ground) is worth to be analysed. In that case, the GLAO system provides a very good correction and PSF are close to be diffraction limited. In that regime, the PSF shape becomes more complex and the impact of inaccurate atmospheric parameters has a more significant impact on the PSFR accuracy. 

Finally let us focus on the last important atmospheric parameter (especially for a large aperture telescope), the outer scale $L_0$. In this case, both $seeing$ and GLF are fixed to their median values (0.8" and 70\%). Four $L_0$ values (8, 16, 24 and 32m) are considered in the simulation in order to span the wide possible range of outer-scale fluctuations. 
\begin{figure}[ht!]
\includegraphics[width=1\linewidth]{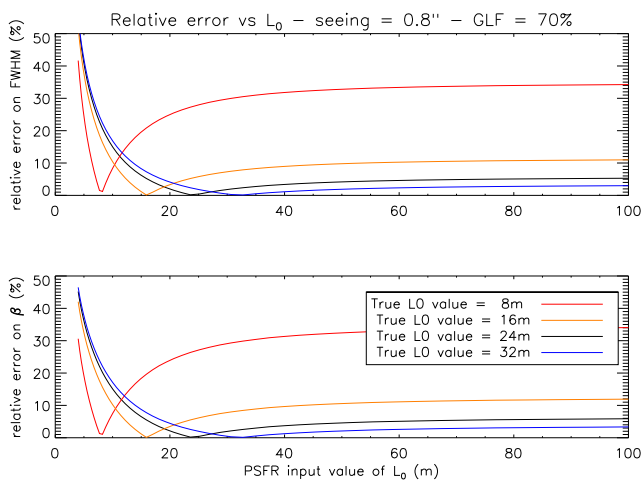}
\caption{Relative error on FWHM and $\beta$ (in \%)  as a function of the bias on $L_0$ estimation. 4 cases of real $L_0$ inputs are considered: 8,16, 24 and 32m. the seeing value is equal to 0.8" and GLF is equal to 70\% (which roughly corresponds to the median values measured on 1 years of MUSE-WFM+GLAO operations).}
\label{fig:psf_accuracy_simul_L0}
\end{figure}
Figure \ref{fig:psf_accuracy_simul_L0} presents the main results obtained for the various real outer-scale values as a function of outer-scale input in the PSFR algorithm. Here, we see that the outer scale estimation is clearly not critical as soon as the real atmospheric outer scale is larger than typically 16m (i.e. two times the telescope diameter). Below this limit, an exact measurement of $L_0$ becomes important. Fortunately, small $L_0$ are rare (from Paranal measurements). More importantly, when the outer scale is small, its signature on the WFS signal becomes relatively strong. Therefore, its estimation from RTC data should be accurate enough assuming that the atmospheric parameter measurement from the RTC data process is properly calibrated and validated. 
The details of these measurements and calibration processes are reported below.  

\section{On-sky data}
The previous sections give an overview of the PSFR algorithm and of its performance measured on simulated data. This algorithm has been implemented in Python (see Section \ref{sec:implementation}) and has been tested and validated on real MUSE-WFM data acquired during commissioning, science verification, and the early operation periods. 
The available data can be split into two main categories: 
\begin{itemize}
\item The RTC (also known as SPARTA) data. The instantaneous WFS measurement and DSM command are used to compute statistics on which turbulence models are fitted \citep{Fusco-2004a,Fusco-2004b,sparta}. For each LGS-WFS signal atmospheric parameters ($r_0$, $L_0$, GLF wind speed) as well as WFS information (e.g. WFS noise) are saved every 30 seconds. An example of a statistical analysis obtained from these data is plotted in Figure \ref{fig:stat_turb} and described in Section \ref{sec:RTC_description}.
\item The MUSE 3D images (of Global Clusters) themselves. MUSE data will be post-processed and Moffat functions will be fitted on them. This fit will produce the `reference values' for our on-sky tests. A detailed description of the MUSE data and their processing is provided in Section \ref{sec:MUSE-data}.
\end{itemize}
\subsection{Adaptive optics telemetry and environmental data}\label{sec:RTC_description}
\begin{figure}[ht!]
\includegraphics[width=1\linewidth]{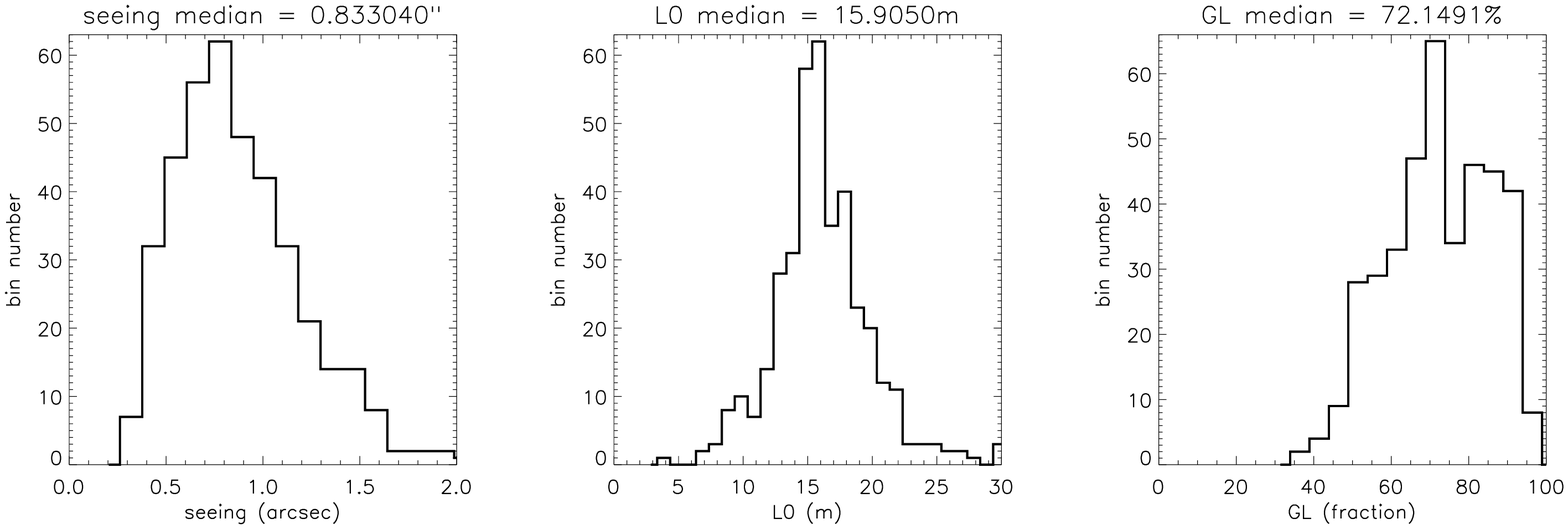} \\
\caption{Real Time Computer DATA: median seeing = 0.83", median $L_0$ = 16m, median GLF = 72\%.}
\label{fig:stat_turb}
\end{figure}

The SPARTA RTC system does continuously (every 30 seconds) provide information on the main atmospheric parameters derived from the LGS WFS and DSM command recorded in closed loop after a pseudo open-loop reconstruction. In the following, we use 392 data sets associated to observations of Globular Clusters performed by MUSE during the periods of  October 1,  2017, to August 31, 2018;   we kept 355 of them. We discarded data with (i) computation issues (aberrant values) and (ii) very large seeing (> 1.5"). 
From the remaining RTC data, turbulence parameters are estimated (in the LOS) and a statistical analysis is provided (see Figure \ref{fig:stat_turb}). The median values of seeing (0.83’’), $L_0$ (16 m) and GLF (72\%)
are fully compatible with the common Paranal values now recorded for more than 20 years. 
The seeing probability density function (PDF) has a typical Poisson distribution shape. The $L_0$ PDF is almost symmetric around its median value with a tight FWHM of only $\pm 3m$. Only a very small percentage of the measurements exhibit an outer scale smaller than 8m and the validity of those data points remains questionable. Finally, the GLF PDF is more structured, with some kind of bi-modal shape, and a significant part of the occurrences are found in the 80-90\% domain. This aspect could be important in the following. The combination of this large GLF with small seeing values should lead to very good GLAO performance and produce near-diffraction limit images (at least at the AO system focal plan output before entering  the MUSE spectrograph). This has two main implications: (i) a higher sensitivity to MUSE internal defects (see Section \ref{sec:MUSE-intern}) and (ii) a more complex final PSF shape than that coming from the Moffat assumption. This latter effect will probably be one of the main limitations of the current method. 

\subsection{MUSE Wide Field Mode 3D data}\label{sec:MUSE-data}
The data were obtained within the MUSE globular cluster survey \citep{Kamann-2018}, which is carried out as part of the MUSE guaranteed time observations. The survey targets the central regions of Galactic globular clusters with a series of relatively short exposures. In order to detect variable stars, the observations of each cluster are split into different epochs, with time lags of hours to months between them. Each individual epoch is split into three exposures, in between which derotator offsets of $90^{\circ}$ are applied. For this work, we consider all the data taken with the WFM AO system between October 1, 2017, and August 31, 2018, that is, in observing periods P100 and P101. In total, 413 individual exposures were analysed.

\subsection{Point-spread-function fit on MUSE data cube}

We performed a fit of the PSF on the MUSE data cube using \textsc{PampelMuse} \citep{2018ascl.soft05021K,2013A&A...549A..71K}, a software package designed for the analysis of integral field data of crowded stellar fields such as globular clusters.  \textsc{PampelMuse} uses a reference catalogue containing the world coordinates and magnitudes of the sources in the observed field as input and first identifies the subset of available sources that can be resolved from the integral field data. In a subsequent step, \textsc{PampelMuse} determines the coordinate transformation from the input catalogue to the integral field data as well as the PSF as a function of wavelength. This information is finally used to optimally extract the spectra of the resolved sources. The MUSE data considered in this study were analysed using an analytical Moffat profile as PSF. Both the width of the Moffat (parametrised by the FWHM) and its kurtosis (parametrised by the parameter $\beta$) were optimised during the analysis and were allowed to change with wavelength.

As mentioned above, a standard observation of a globular cluster consists of three exposures with derotator offsets. By default, the exposures are combined before the analysis, in order to homogenise the image quality across the FOV. However, for this project we analysed the individual exposures, which allows for a more direct comparison with the atmospheric parameters gathered during the observations. As the resampling into a data cube can produce artefacts if only a single exposure is used, \textsc{PampelMuse} has been modified to work on pixel tables, an intermediate data format used by the MUSE pipeline that does not require resampling \citep[see][]{2014ASPC..485..451W}.

\subsection{Results}
In order to investigate the quality of the PSF fits, we proceeded as follows. When analysing an exposure, \textsc{PampelMuse} selects a number of bright and reasonably isolated stars that are used to optimise the PSF model. The optimisation is done iteratively. The contributions of nearby stars that could potentially disturb the fits are subtracted using an initial PSF model, after which the model is refined by fitting single PSF profiles to the selected stars. The refined model is then used to improve the subtraction of the nearby stars. The steps are repeated until the fluxes of the PSF stars have converged.

\begin{figure}
\includegraphics[width=1\linewidth]{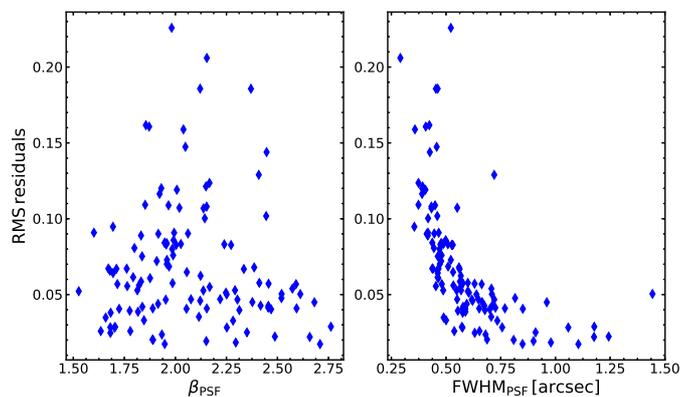}
\caption{Average relative residuals of the PSF fits in the central $2\arcsec$ as a function of $\beta$  ({\it left}) and the FWHM ({\it right}) of the Moffat profile used to fit the MUSE WFM-AO PSF. 
}
\label{fig:pampelmuse_psf_accuracy}
\end{figure}

After convergence, we extracted radial profiles of the PSF stars and compared them to the radial profiles of the models. We note that before extracting the profiles from the integral field data, we again subtracted the contributions of nearby stars. By subtracting the model profile from the measured one and dividing the result by the measured profile, we determined  the relative residuals for each PSF fit. Those were averaged for the 50\% brightest PSF stars. Finally, we measured the RMS deviation from zero of the mean relative residuals within the central $2\arcsec$. This value, which is shown as a function of $\beta$ and the FWHM of the fitted Moffat PSF in Fig.~\ref{fig:pampelmuse_psf_accuracy}, serves as our criterion for the agreement between our PSF model and the actual MUSE WFM-AO PSF. It can be understood as the typical residual flux in a pixel after subtraction of a star relative to its recorded flux.
The results depicted in Fig.~\ref{fig:pampelmuse_psf_accuracy} show that the residuals of the PSF fits are typically $<10\%$, although some cases exist where the Moffat profiles seem to provide a less accurate fit to the actual PSF. While no obvious trend with the fitted values of $\beta$ is visible, there is an anti-correlation between the strength of the residuals and the value of the fitted FWHM. While the fit residuals are typically $<5\%$ for observations with ${\rm FWHM} > 0.6\arcsec$, stronger residuals are observed for smaller FWHM values. We attribute this behaviour to the PSF becoming critically sampled. The spatial sampling of MUSE in the WFM is $0.2\arcsec$, meaning that a PSF with a FWHM of $0.4\arcsec$ will be approximately Nyquist sampled. Hence, as the width of the PSF approaches this limit,  it becomes increasingly difficult to recover its true shape. A direct consequence of this observation is that one has to expect larger PSF residuals for data obtained under better conditions.

\section{On-sky performance of the PSF reconstruction}
Using both the PSFR estimate computed with the RTC data and the associated results of the PSF fit on the Globular Cluster images, we can now test and assess the performance of our algorithm on real on-sky data. 
\subsection{Point-spread-function\ reconstruction and first comparison with MUSE data}
The Fourier algorithm provides us with a GLAO PSF but does not account for any instrumental defects. The MUSE image quality is mostly dominated by its 0\farcs2 sampling. Measurements performed during the `Preliminary Acceptance in Europe' indicate an image quality (FWHM) of between 0.20\farcs \ and 0.27\farcs, depending of the channel, and over the full wavelength range. Without any additional available information, the most straightforward way to include that instrumental defects is to convolve the GLAO PSF with a Gaussian function of 0.2\farcs  FWHM. 
\begin{figure*}[ht!]
\includegraphics[width=1\linewidth]{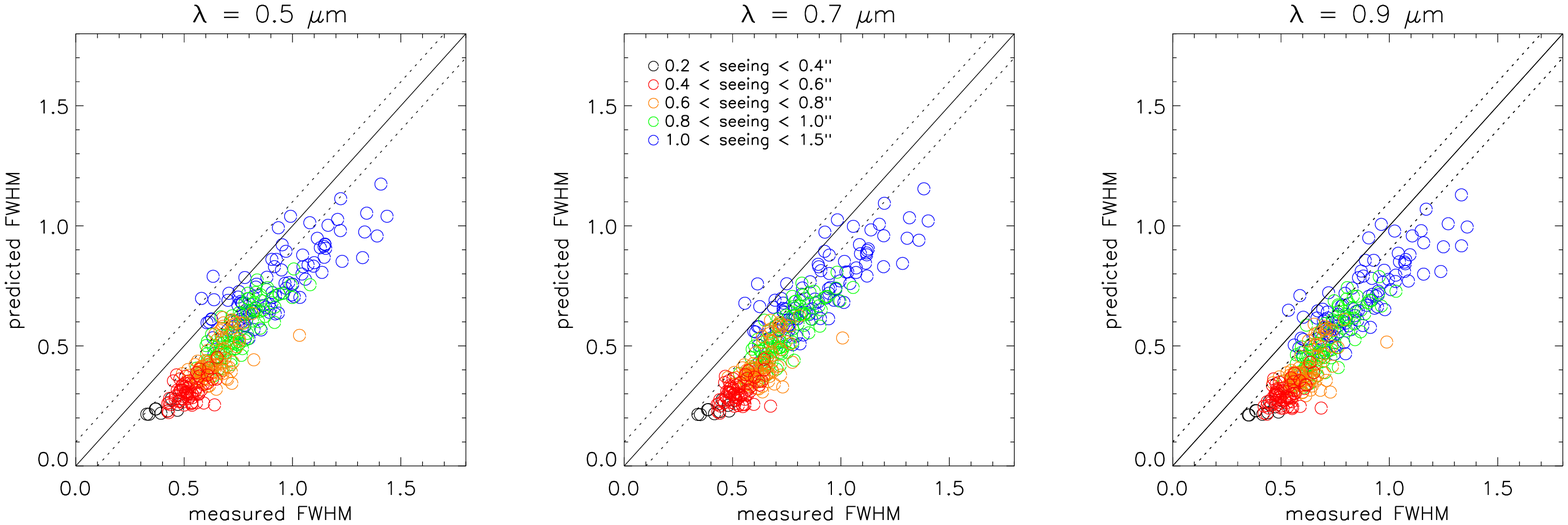}
\ \\
\includegraphics[width=1\linewidth]{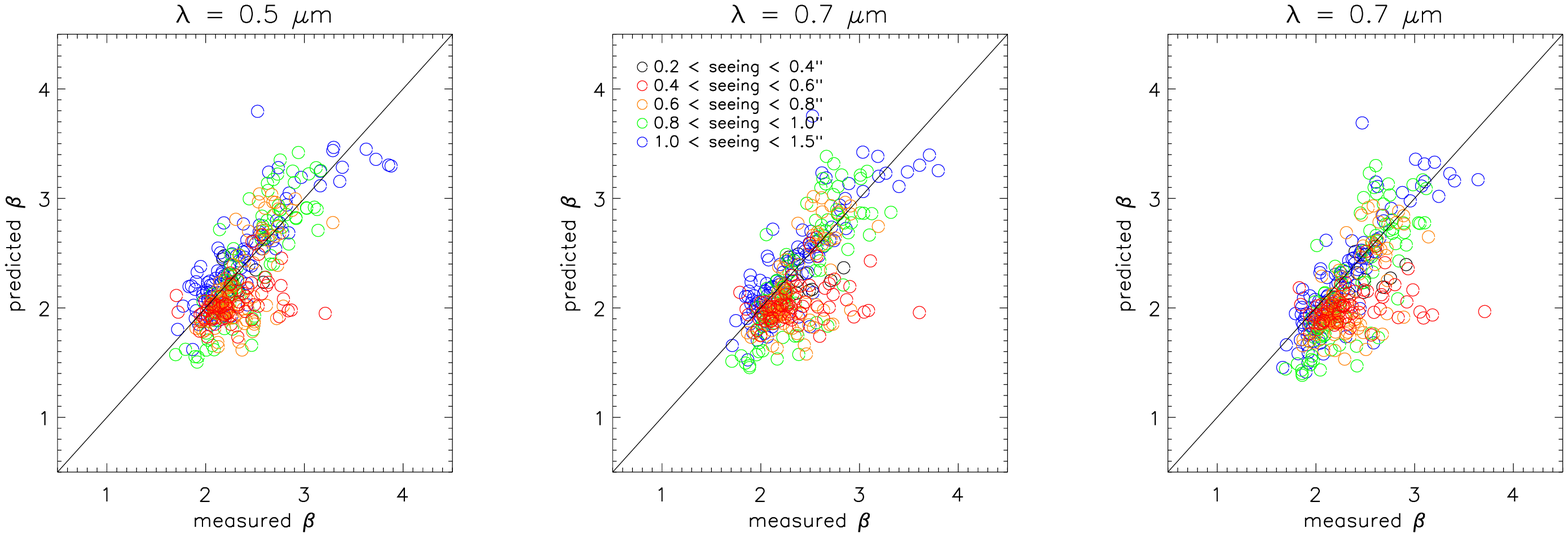}
\caption{Comparison of FWHM (top) and $\beta$ parameter (bottom) for measured data on MUSE images (extract from Globular Custer data) and PSFR data (computed from atmospheric data obtained using AO telemetry) convolved by a 0.2" FWHM Gaussian function (to account for MUSE pixel size). The various colors stand for various seeing values domain (black:0.2-0.4" -- red:0.4-0.6" -- orange:0.6-0.8" -- green:0.8-1.0" -- blue:1.0-1.5"\label{fig:psfR-on-sky-pixelonly}}
\end{figure*}
Even though a very good correlation (defined as a classical Pearson correlation coefficient) 
is visible in Figure \ref{fig:psfR-on-sky-pixelonly} (more than 90\% for FWHM and more than 70\% for $\beta$), a bias characterised by an underestimation of the FWHM is clearly visible with a typical value of 0.25 and 0.3", respectively.

\subsection{Non-atmospheric part to the PSF}\label{sec:MUSE-intern}
\begin{figure}
\includegraphics[width=1.\linewidth]{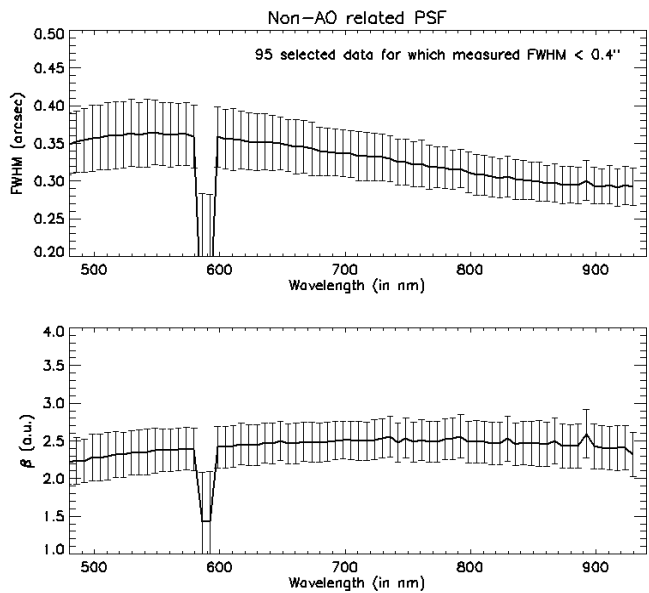}
\caption{Full width at half maximum and $\beta$ parameter estimated for the MUSE internal PSF. A set of FWHM and $\beta$  is estimated for each wavelength bin between 480 and 940 nm. The blind area around 589nm is determined by the notch filter which blocks the laser guide star light in the instrument.}
\label{fig:MUSE-internal-PSF}
\end{figure}

The bias observed in the previous section is mainly attributed to the MUSE internal PSF.
The very nature of the instrument makes a precise measurement of the internal PSF, both in the
FoV and for each wavelength
channel, very challenging (if not impossible). No such measurement was available with a sufficient spatial and spectral resolution. 
To deal with this specific issue we decided to measure the overall MUSE internal PSF (including both the optics and the detector) using on-sky data. To do so, the following multi-step process was applied: 
\begin{itemize}
\item Identification of a subset of data points (among the 355 available). We chose the best data points of the data set (those for which, at the reddest wavelength,  the PSF FWHM computed on the MUSE images is better than 0.4 arcsec). This corresponds to 95 of the 392 data points, that is typically 27\% of the data. 
\item On this data subset, for each MUSE wavelength, the GLAO PSFR computed with our algorithm were convolved by a MUSE internal PSF ($PSF_{MUSE}$ also modelled by a Moffat function characterised by its $FWHM_{MUSE}(\lambda)$ and $\beta_{MUSE}(\lambda)$ parameters). Using a classical least square error metric we adjust $FWHM_{MUSE}(\lambda)$ and $\beta_{MUSE}(\lambda)$  in order to globally minimise the quadratic distance between the subset of MUSE on-sky data and the associated PSFR convolved with $PSF_{MUSE}$, that is, for each wavelength $\lambda$ and each data point of the subset:
\end{itemize}
\begin{eqnarray}
  & \text{Min}\left\{ \left|\left| PSF_{meas,i} - PSFR_{i,\lambda}^{FWHM,\beta} \right|\right|^2 \right\} \\ \nonumber
 & \text{with respect to }FWHM \text{ and } \beta,
\end{eqnarray}
with 
\begin{equation}
\begin{aligned}
&PSFR_{i,\lambda}^{FWHM,\beta} = \\
&PSF_{Tel}\star PSF_{GLAO}(\lambda_i) \star M(FWHM_\text{MUSE}(\lambda_i),\beta_\text{MUSE}(\lambda_i)).\\
\end{aligned}
\end{equation}
This process allows us to obtain an associated MUSE internal PSF characterised by its FWHM and $\beta$ parameter for each MUSE wavelength channel. This PSF includes both detectors and optical defects. The final results are plotted in Figure \ref{fig:MUSE-internal-PSF}. 
The results are fully compatible with the MUSE original specification and with the very few and incomplete MUSE PSF internal measurements made in the laboratory during the Assembly Integration and Test (AIT) period. The MUSE internal PSF (detector included) goes from 0.35" to 0.3" (in the reddest part of the instrument spectrum). Assuming a 0.2" detector FWHM, this corresponds to a full optical error budget of between 0.25 and 0.2". 

We note that we assumed here that the MUSE internal aberrations were, are, and will be fully static temporally speaking. By design of the instrument and because the instrument lays on the VLT Nasmyth platform we strongly believe that this assumption will remain correct at the level of accuracy required for the MUSE PSF reconstruction algorithm (i.e. that any temporally variable internal aberrations will not affect the internal PSF by more than a few tens of mas in terms of FWHM).  
 Considering that level of GLAO correction and the level of required accuracy on the reconstructed PSF, an error on the static aberration of a few tens (up to a few hundreds) of nanometers will be completely negligible compared to the atmospheric contribution. This is far larger than any expected temporal evolution of the instrumental aberrations.
The comparison between our measurements and some partial (for only a very limited number of the MUSE channels) internal data acquired during the AIT stage of the instrument seems to fully confirm this hypothesis. In any case, a follow up of the internal aberrations with time could be organised using the same procedure proposed here in order to fully validate the hypothesis.   

The MUSE PSF is now included in the complete PSFR algorithm in order to obtain the final performance. Although it has been computed using only the best available data, it is now applied to all the data, assuming that this MUSE internal PSF remains constant during the lifetime of the project (or at least between two re-calibration process).

\subsection{Final performance}
\begin{figure*}[ht!]
\includegraphics[width=1\linewidth]{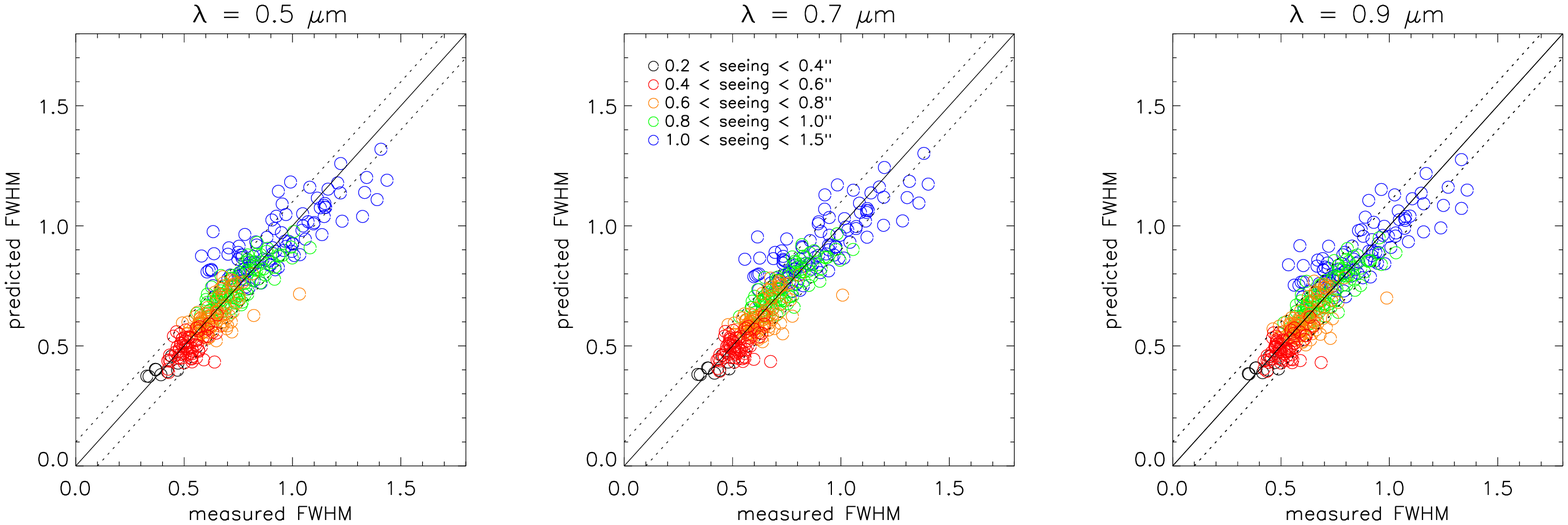}
\ \\
\includegraphics[width=1\linewidth]{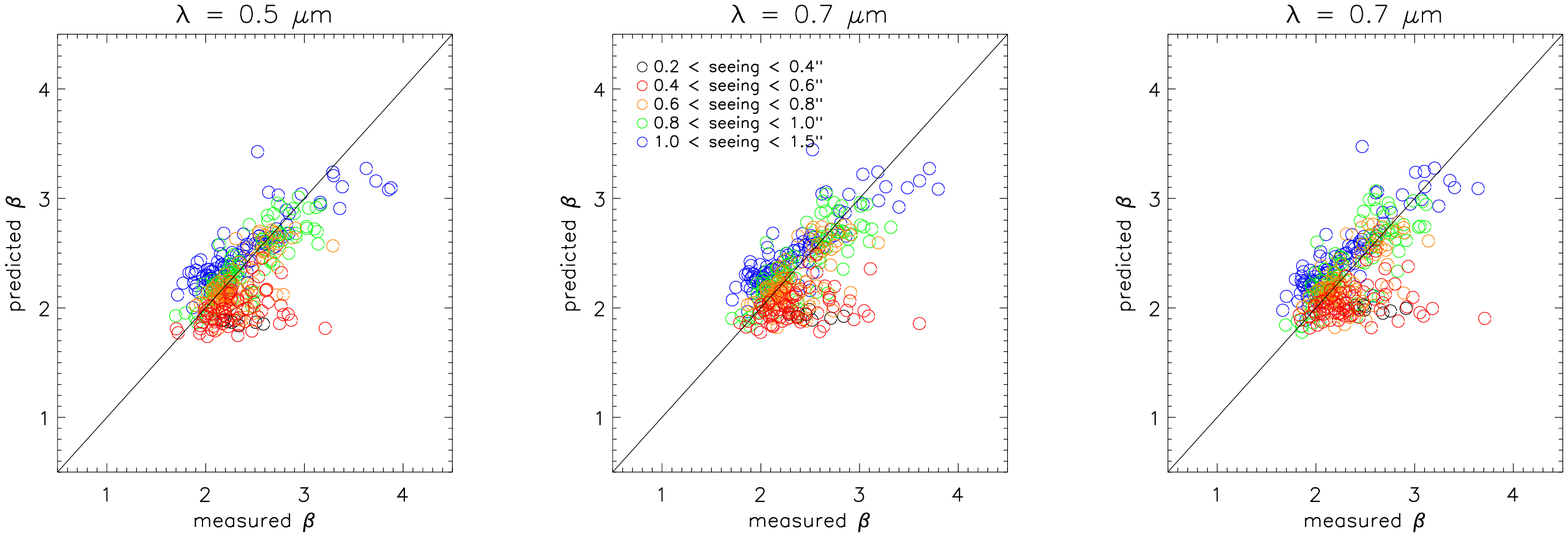}
\caption{As in Figure \ref{fig:psfR-on-sky-pixelonly}, but  the pixel PSF has been replaced by a full internal MUSE PSF.}
\label{fig:psf_accuracy}
\end{figure*}

Let us now use the MUSE internal PSF in the full PSFR process and re-process all the data (the 355 available) with the final version of the algorithm. The results are plotted in Figure \ref{fig:psf_accuracy}.
Results can be compared to Figure \ref{fig:psfR-on-sky-pixelonly}. The correlation of both FWHM and $\beta$ remains identical but the bias has completely disappeared for ALL the processed data, showing the pertinence of the MUSE internal PSF for the whole set of data. 
In order to obtain more quantitative results, we propose in the following to compute an error metric between the final computed parameters using the full PSFR process (including the MUSE internal PSF) and the measured parameters obtained on the Globular Cluster images: 
\begin{equation}
err_{p} = p_{measured} - p_{predicted}
.\end{equation}
Figure \ref{fig:psf_accuracy} shows the PDF of the $err_p$ for both FWHM and $\beta$ for several wavelength ranges. It also shows the PDF cumulative function in each case. 
\begin{figure*}[ht!]
\includegraphics[width=1.\linewidth]{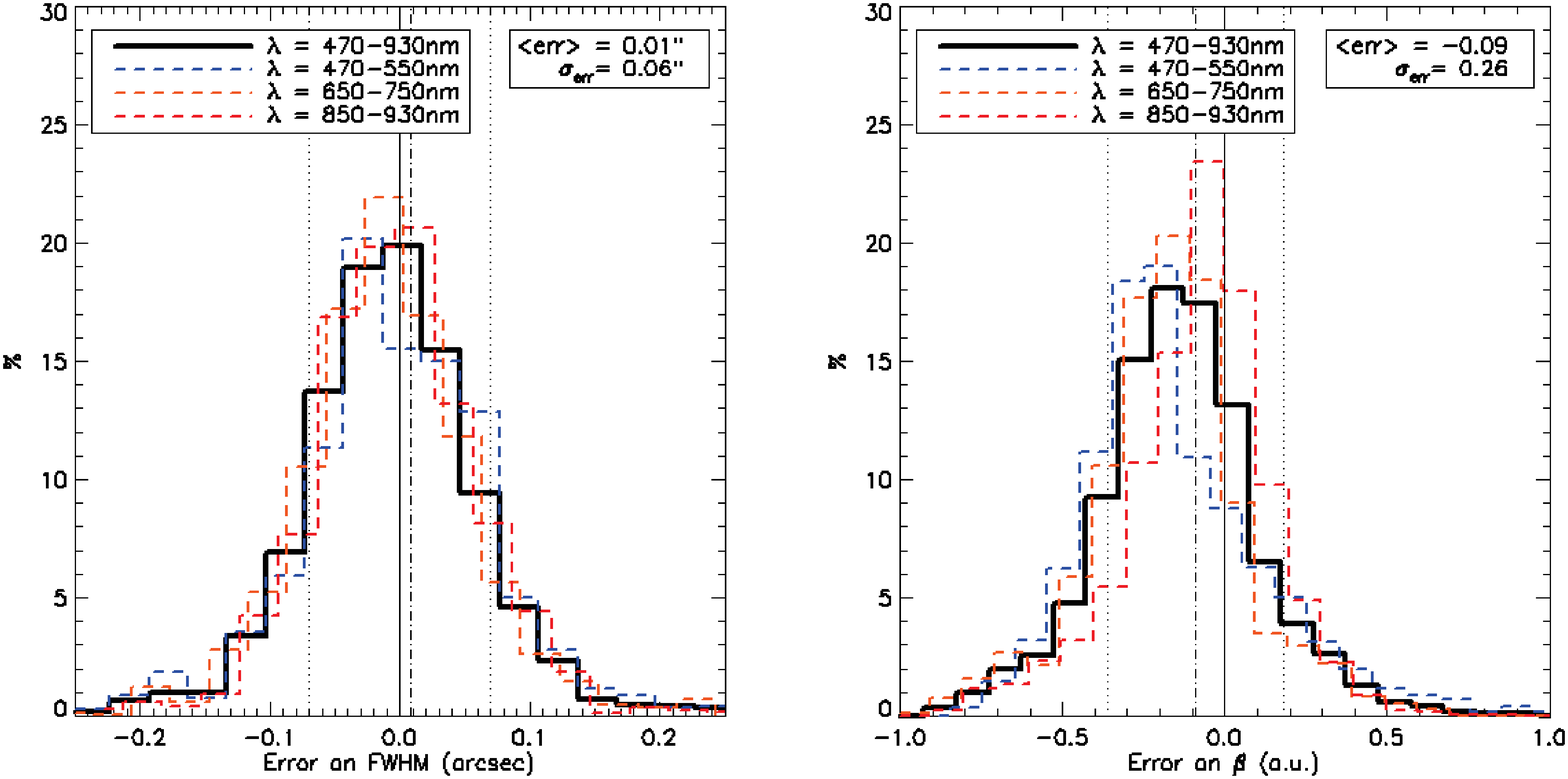}
\ \\
\includegraphics[width=1.\linewidth]{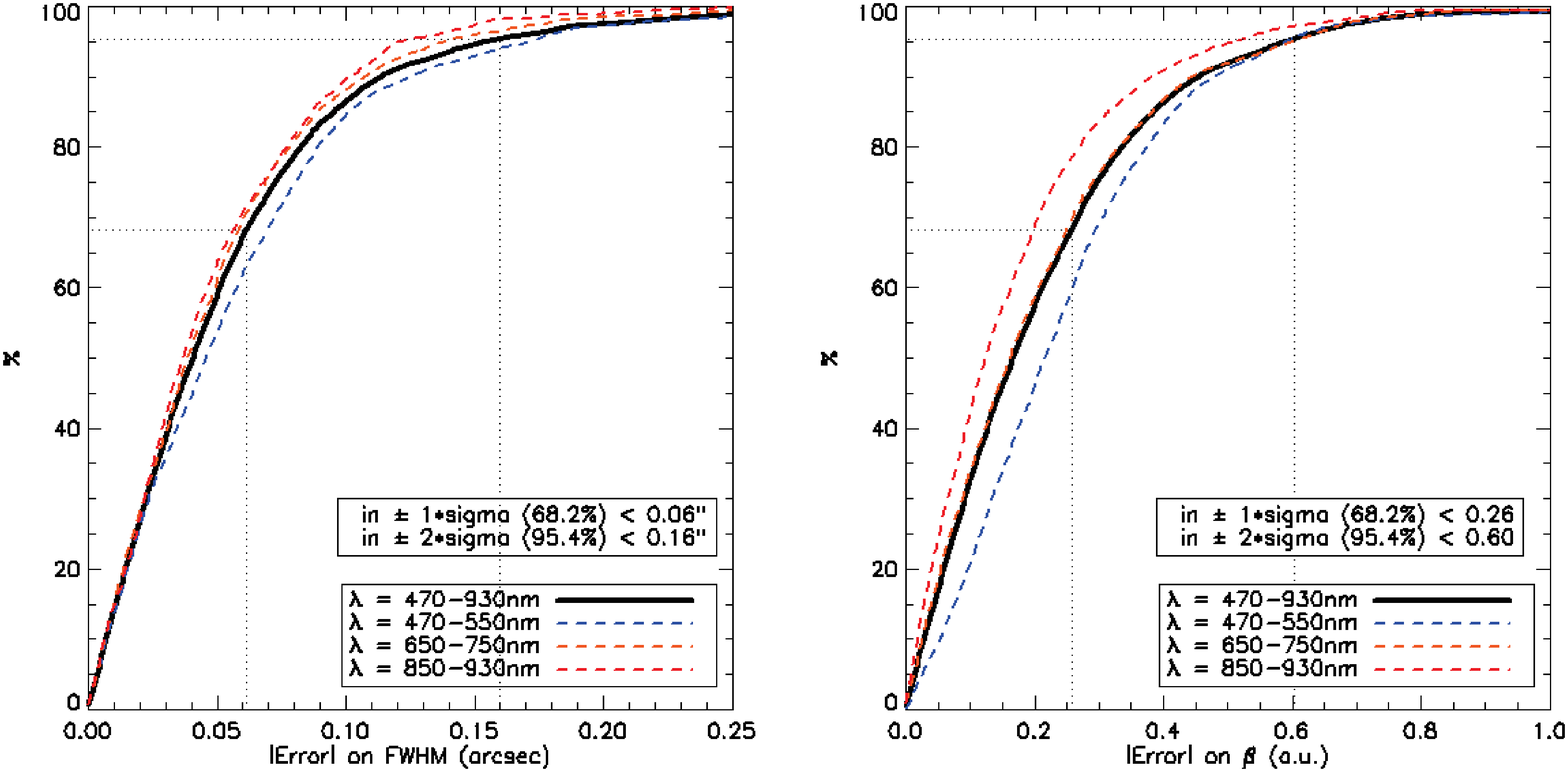}
\caption{Probablity density function of the error on Moffat paremeters (FWHM and $\beta$) for several wavelength ranges.}
\label{fig:psf_accuracy1}
\end{figure*}
From these various plots, we can extract the final on-sky performance of our PSFR algorithm: 
\begin{itemize}
    \item For FWHM,  the bias is 10mas and the 1 and 2 $\sigma$ dispersion are respectively 60 mas and 160 mas (which means that the error is smaller than 60 mas in 68\% of the cases and smaller than 160 mas in 95.4\% of the cases). 
    \item For $\beta,$  the bias is -0.1 and the 1 and 2 $\sigma$ dispersions are respectively 0.26 and 0.6  (which means that the error is smaller than 0.26 in 68\% of the cases and smaller than 0.6  in 95.4\% of the cases). 
\end{itemize}

These results, combined with the simulation analysis, fully demonstrate (with respect to our initial scientific requirements) the accuracy and reliability of the PSFR algorithm. This study validates the proposed strategy and allow us to pass to the next level of the project: the final implementation in the MUSE-WFM pipeline and the use of PSFR for scientific observations and final astrophysical data processing. 

\section{Implementation}\label{sec:implementation}

The PSF reconstruction algorithm is implemented as a Python package (\texttt{muse\_psfr}), and its source code is available on GitHub\footnote{https://github.com/musevlt/muse-psfr}.

The algorithm requires three values provided by SPARTA (the AOF Real Time Computer \citealt{sparta}): the seeing, the ground layer fraction (GLF), and the outer-scale ($L_0$). These values can be provided directly as command-line arguments:

\begin{verbatim}
$ muse-psfr --no-color --values 1,0.7,25
MUSE-PSFR version 1.0rc2
Computing PSF Reconstruction from Sparta data
Processing SPARTA table with 1 values, njobs=1
Compute PSF with seeing=1.00 GL=0.70 L0=25.00
---------------------------------------------
LBDA 5000 7000 9000
FWHM 0.85 0.73 0.62
BETA 2.73 2.55 2.23
---------------------------------------------
\end{verbatim}

It is also possible to provide a raw MUSE file. Since the GLAO commissioning, the MUSE raw files contain a FITS table (\texttt{SPARTA\_ATM\_DATA}) containing the atmospheric turbulence profile estimated by SPARTA. This table contains the values for each laser, with one row every two minutes.

\begin{verbatim}
$ muse-psfr MUSE.2018-08-13T07:14:11.128.fits.fz
MUSE-PSFR version 0.31
OB MXDF-01-00-A 2018-08-13T07:39:21.835 
Airmass 1.49-1.35
Computing PSF Reconstruction from Sparta data
Processing SPARTA table with 13 values, njobs=-1
4/13 : Using only 3 values out of 4 ...
4/13 : seeing=0.57 GL=0.75 L0=18.32
Using three lasers mode
1/13 : Using only 3 values out of 4 ...
1/13 : seeing=0.71 GL=0.68 L0=13.60
Using three lasers mode
6/13 : Using only 3 values out of 4 ...
6/13 : seeing=0.60 GL=0.75 L0=16.47
Using three lasers mode
....

OB MXDF-01-00-A 2018-08-13T07:39:21.835 
Airmass 1.49-1.35
---------------------------------------
LBDA  5000 7000 9000
FWHM  0.57 0.46 0.35
BETA  2.36 1.91 1.64
---------------------------------------
\end{verbatim}

The last option is to use the Python API directly, which gives access to more parameters:
\begin{itemize}
    \item \emph{Number of reconstructed wavelengths:} To reduce computation time, the \texttt{muse-psfr} command reconstructs the PSF at three wavelengths: 500, 700, and 900 nm. But it is possible to reconstruct the PSF at any wavelength, with the \texttt{compute\_psf\_from\_sparta} function. This function reconstructs by default for 35 wavelengths between 490nm and 930nm (which can be specified with the \texttt{lmin}, \texttt{lmax}, and \texttt{nl} parameters)
    \item \emph{Number of reconstructed directions:} Since the spatial variation is negligible over the MUSE FOV, the reconstruction is done by default only at the centre of the FOV. This can be changed in \texttt{compute\_psf\_from\_sparta} with the \texttt{npsflin} parameter.
\end{itemize}

The documentation\footnote{https://muse-psfr.readthedocs.io/} gives more information about the Python API and the various parameters.

\section{Example of application: MUSE deep field}
\begin{figure}[ht!]
\includegraphics[width=1\linewidth]{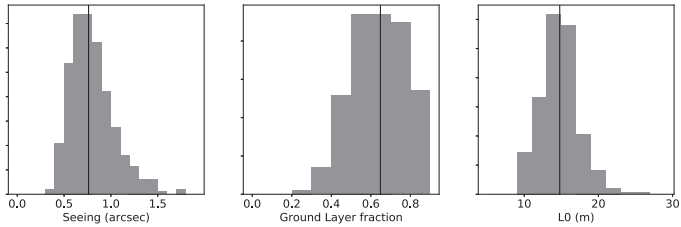} 
\caption{Histogram of atmospheric parameters measured by the SPARTA real-time AO controller during the MXDF observations. The solid line displays the median value.}
\label{fig:mxdf-atm}
\end{figure}

\begin{figure}[ht!]
\includegraphics[width=1\linewidth]{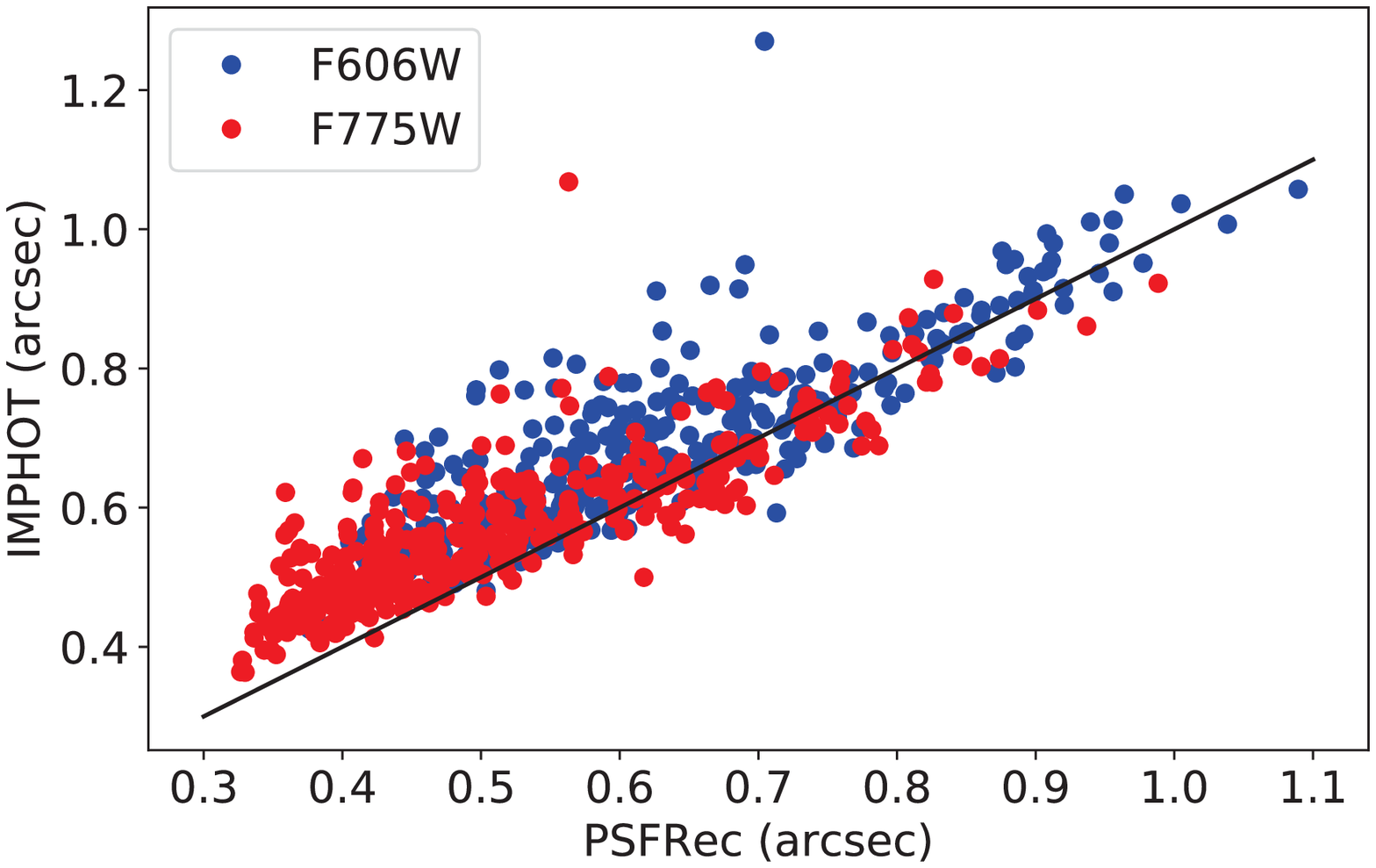} 
\caption{Comparison between Reconstructed PSF Moffat FWHM (PSFRec) and values derived from comparison with high-resolution HST broad-band images (IMPHOT) for two filters (F606W and F775W).}
\label{fig:mxdf}
\end{figure}

We used the algorithm to estimate the PSF for the MUSE eXtreme Deep fields (MXDF) obervations performed with MUSE in the area of the Hubble Ultra Deep Field \citep{Beckwith2006}. The aim of the project (Bacon et al in prep) is to perform the deepest ever spectroscopic deep field by accumulating more than 100 hours of integration in a single MUSE field. The observations were performed between August 2018 and February 2019 with the GLAO mode. The field location was selected to be in the deepest region of the UDF and to have a tip/tilt star bright enough to ensure a good GLAO correction, plus a fainter star in the outskirts of the MUSE FOV for the slow guiding system. These two requirements allow us to achieve the best possible spatial resolution for the given atmospheric conditions. However, given the poor star density at this location, it was not possible to simultaneously have  a PSF star in the FOV and thus an alternative way to estimate the PSF was required.

A total of 377 exposures with 25 mn integration time was obtained. As shown in Fig.~\ref{fig:mxdf-atm}, exposures were taken in a variety of seeing and ground-layer conditions. For each exposure, we compute with \texttt{muse-psfr} a polynomial approximation of the Moffat parameters $FWHM(\lambda)$ and $\beta(\lambda)$. 

We also used the \texttt{imphot} method to estimate the PSF. The method uses the high-resolution HST broad-band images with the corresponding broad-band MUSE reconstructed images to derive the convolution Moffat kernel which minimises the difference between the two images. The method is described in detail in \cite{Bacon2017}. We use two HST broad-band images in the F606W and F775W filters which cover the MUSE wavelength range. The corresponding \texttt{muse-psfr} FWHM value at the filter central wavelength is shown for comparison in Fig.~\ref{fig:mxdf}. The two methods are in good agreement with a scatter of 0.06 arcsec rms for both filters. We note that a systematic mean offset of 0.06 arcsec is measured between the two methods, the \texttt{imphot} method giving higher FWHM than \texttt{muse-psfr} when the PSF is small. This bias is most likely due to the way the sampling is taken into account but it is difficult to come to any conclusions on this matter without an independent ground truth measurement.

\section{Conclusions}
We present a simple, efficient, and fully operational (from the astrophysical image analysis of MUSE WFM images) PSF-reconstruction algorithm based on a Fourier analysis of the GLAO residual phase statistics completed by dedicated information and measurements concerning the instrument itself. A detailed analysis of the GLAO error budget has allowed us to both simplify and optimise the algorithm. It has been thoroughly and successfully tested with respect to complete End-to-End simulations. A sensitivity analysis allowed us to determine the required accuracy in terms of input parameters. It is shown that precise knowledge (typically a few percent accuracy) on seeing, ground layer, and outer-scale values is required to fulfill the astronomer requirements. The code was then tested with respect to real on-sky data obtained during the commissioning and the science verification of the coupling of MUSE-WFM and the AOF. Gathering almost 400 independent observations of globular clusters, PSF reconstruction was performed for each set of data and a statistical analysis of the results was performed. Using a subsample of the data (only those obtained under the best observation conditions) it has been possible to estimate the MUSE internal PSF (at each wavelength). After integration of the MUSE internal PSF into the algorithm, we demonstrated that it is now capable of reconstructing the critical parameters of a PSF (represented by the FWHM and the kurtosis parameters of the Moffat function) with the accuracy required by astronomers in 90\% of the observing cases. More precisely, we obtain an error on the PSF FWHM of smaller than 60 mas (less than one-third of a pixel) for 68\% of the cases and 160 mas (smaller than the pixel) for 95.4\% of the cases. Similarly, for the $\beta$ parameter, an error smaller than 0.26 is obtained for 68\% of the cases and smaller than 0.6 for 95.4\%. 
After this successful validation, the algorithm was implemented as a python package and can be now used routinely with the MUSE 3D data. A first example of application is presented here for the MUSE deep field observations. It is now a fully operational tool available for all  users of MUSE. 

\begin{acknowledgements}
This work has been partially supported by the ANR-APPLY program (ANR-19-CE31-0011), the European Research Council consolidator grant (ERC-CoG-646928-Multi-Pop) and by OPTICON, in the Horizon 2020 Framework Program of the European Commission's (Grant number 730890 - OPTICON).
\end{acknowledgements}


\begin{thebibliography}{37}
\expandafter\ifx\csname natexlab\endcsname\relax\def\natexlab#1{#1}\fi

\bibitem[{{Andersen} {et~al.}(2006){Andersen}, {Stoesz}, {Morris},
  {Lloyd-Hart}, {Crampton}, {Butterley}, {Ellerbroek}, {Jolissaint}, {Milton},
  {Myers}, {Szeto}, {Tokovinin}, {V{\'e}ran}, \& {Wilson}}]{Andersen2006}
{Andersen}, D.~R., {Stoesz}, J., {Morris}, S., {et~al.} 2006, \pasp, 118, 1574

\bibitem[{Arsenault {et~al.}(2008)Arsenault, Madec, Hubin, Paufique, Stroebele,
  Soenke, Donaldson, Fedrigo, Oberti, Tordo, Downing, Kiekebusch, Conzelmann,
  Duchateau, Jost, Hackenberg, Calia, Delabre, Stuik, Biasi, Gallieni,
  Lazzarini, Lelouarn, \& Glindeman}]{Arsenault-2008}
Arsenault, R., Madec, P.-Y., Hubin, N., {et~al.} 2008, in Adaptive Optics
  Systems, ed. N.~Hubin, C.~E. Max, \& P.~L. Wizinowich, Vol. 7015,
  International Society for Optics and Photonics (SPIE), 577 -- 588

\bibitem[{{Bacon} {et~al.}(2004){Bacon}, {Bauer}, {Bower}, {Cabrit},
  {Cappellari}, {Carollo}, {Combes}, {Davies}, {Delabre}, {Dekker},
  {Devriendt}, {Djidel}, {Duchateau}, {Dubois}, {Emsellem}, {Ferruit}, {Franx},
  {Gilmore}, {Guiderdoni}, {Henault}, {Hubin}, {Jungwiert}, {Kelz}, {Le
  Louarn}, {Lewis}, {Lizon}, {McDermid}, {Morris}, {Laux}, {Le F{\`e}vre},
  {Lantz}, {Lilly}, {Lynn}, {Pasquini}, {Pecontal}, {Pinet}, {Popovic},
  {Quirrenbach}, {Reiss}, {Roth}, {Steinmetz}, {Stuik}, {Wisotzki}, \& {de
  Zeeuw}}]{Bacon2004}
{Bacon}, R., {Bauer}, S.-M., {Bower}, R., {et~al.} 2004, Society of
  Photo-Optical Instrumentation Engineers (SPIE) Conference Series, Vol. 5492,
  {The second-generation VLT instrument MUSE: science drivers and instrument
  design}, ed. A.~F.~M. {Moorwood} \& M.~{Iye}, 1145--1149

\bibitem[{{Bacon} {et~al.}(2017){Bacon}, {Conseil}, {Mary}, {Brinchmann},
  {Shepherd}, {Akhlaghi}, {Weilbacher}, {Piqueras}, {Wisotzki}, {Lagattuta},
  {Epinat}, {Guerou}, {Inami}, {Cantalupo}, {Courbot}, {Contini}, {Richard},
  {Maseda}, {Bouwens}, {Bouch{\'e}}, {Kollatschny}, {Schaye}, {Marino},
  {Pello}, {Herenz}, {Guiderdoni}, \& {Carollo}}]{Bacon2017}
{Bacon}, R., {Conseil}, S., {Mary}, D., {et~al.} 2017, \aap, 608, A1

\bibitem[{{Bacon} {et~al.}(2014{\natexlab{a}}){Bacon}, {Vernet}, {Borisova},
  {Bouch{\'e}}, {Brinchmann}, {Carollo}, {Carton}, {Caruana}, {Cerda},
  {Contini}, {Franx}, {Girard}, {Guerou}, {Haddad}, {Hau}, {Herenz}, {Herrera},
  {Husemann}, {Husser}, {Jarno}, {Kamann}, {Krajnovic}, {Lilly}, {Mainieri},
  {Martinsson}, {Palsa}, {Patricio}, {P{\'e}contal}, {Pello}, {Piqueras},
  {Richard}, {Sandin}, {Schroetter}, {Selman}, {Shirazi}, {Smette}, {Soto},
  {Streicher}, {Urrutia}, {Weilbacher}, {Wisotzki}, \& {Zins}}]{Bacon2014}
{Bacon}, R., {Vernet}, J., {Borisova}, E., {et~al.} 2014{\natexlab{a}}, The
  Messenger, 157, 13

\bibitem[{{Bacon} {et~al.}(2014{\natexlab{b}}){Bacon}, {Vernet}, {Borisova},
  {Bouch{\'e}}, {Brinchmann}, {Carollo}, {Carton}, {Caruana}, {Cerda},
  {Contini}, {Franx}, {Girard}, {Guerou}, {Haddad}, {Hau}, {Herenz}, {Herrera},
  {Husemann}, {Husser}, {Jarno}, {Kamann}, {Krajnovic}, {Lilly}, {Mainieri},
  {Martinsson}, {Palsa}, {Patricio}, {P{\'e}contal}, {Pello}, {Piqueras},
  {Richard}, {Sandin}, {Schroetter}, {Selman}, {Shirazi}, {Smette}, {Soto},
  {Streicher}, {Urrutia}, {Weilbacher}, {Wisotzki}, \& {Zins}}]{Bacon-2014}
{Bacon}, R., {Vernet}, J., {Borisova}, E., {et~al.} 2014{\natexlab{b}}, The
  Messenger, 157, 13

\bibitem[{{Beckwith} {et~al.}(2006){Beckwith}, {Stiavelli}, {Koekemoer},
  {Caldwell}, {Ferguson}, {Hook}, {Lucas}, {Bergeron}, {Corbin}, {Jogee},
  {Panagia}, {Robberto}, {Royle}, {Somerville}, \& {Sosey}}]{Beckwith2006}
{Beckwith}, S. V.~W., {Stiavelli}, M., {Koekemoer}, A.~M., {et~al.} 2006, \aj,
  132, 1729

\bibitem[{Beltramo-Martin {et~al.}(2019)Beltramo-Martin, Correia, Ragland,
  Jolissaint, Neichel, Fusco, \& Wizinowich}]{Beltramo-Martin-2019}
Beltramo-Martin, O., Correia, C.~M., Ragland, S., {et~al.} 2019, Monthly
  Notices of the Royal Astronomical Society, 487, 5450

\bibitem[{{Bendinelli} {et~al.}(1987){Bendinelli}, {Zavatti}, \&
  {Parmeggiani}}]{Bendinelli1987}
{Bendinelli}, O., {Zavatti}, F., \& {Parmeggiani}, G. 1987, Journal of
  Astrophysics and Astronomy, 8, 343

\bibitem[{Bouch{\'e} {et~al.}(2015)Bouch{\'e}, Carfantan, Schroetter,
  Michel-Dansac, \& Contini}]{bouche-2015}
Bouch{\'e}, N., Carfantan, H., Schroetter, I., Michel-Dansac, L., \& Contini,
  T. 2015, The Astronomical Journal, 150, 92

\bibitem[{Conan \& Correia(2014)}]{OOMAO}
Conan, R. \& Correia, C. 2014, in Adaptive Optics Systems IV, ed. E.~Marchetti,
  L.~M. Close, \& J.-P. Véran, Vol. 9148, International Society for Optics and
  Photonics (SPIE), 2066 -- 2082

\bibitem[{Damjanov {et~al.}(2011)Damjanov, Abraham, Glazebrook, McGregor,
  Rigaut, McCarthy, Brinchmann, Cuillandre, Mellier, McCracken,
  {et~al.}}]{damjanov-2011}
Damjanov, I., Abraham, R.~G., Glazebrook, K., {et~al.} 2011, Publications of
  the Astronomical Society of the Pacific, 123, 348

\bibitem[{Epinat {et~al.}(2010)Epinat, Amram, Balkowski, \&
  Marcelin}]{epinat-2010}
Epinat, B., Amram, P., Balkowski, C., \& Marcelin, M. 2010, Monthly Notices of
  the Royal Astronomical Society, 401, 2113

\bibitem[{Fedrigo {et~al.}(2006)Fedrigo, Donaldson, Soenke, Goodsell, Geng,
  Saunter, \& Dipper}]{sparta}
Fedrigo, E., Donaldson, R., Soenke, C., {et~al.} 2006, Proc SPIE

\bibitem[{{F{\'e}tick} {et~al.}(2019){F{\'e}tick}, {Fusco}, {Neichel},
  {Mugnier}, {Beltramo-Martin}, {Bonnefois}, {Petit}, {Milli}, {Vernet},
  {Oberti}, \& {Bacon}}]{Fetick-2019b}
{F{\'e}tick}, R.~J.~L., {Fusco}, T., {Neichel}, B., {et~al.} 2019, \aap, 628,
  A99

\bibitem[{{F\'etick, R. JL.} {et~al.}(2019){F\'etick, R. JL.}, {Jorda, L.},
  {Vernazza, P.}, {Marsset, M.}, {Drouard, A.}, {Fusco, T.}, {Carry, B.},
  {Marchis, F.}, {Hanus, J.}, {Viikinkoski, M.}, {Birlan, M.}, {Bartczak, P.},
  {Berthier, J.}, {Castillo-Rogez, J.}, {Cipriani, F.}, {Colas, F.},
  {Dudzi\'{}nski, G.}, {Dumas, C.}, {Ferrais, M.}, {Jehin, E.}, {Kaasalainen,
  M.}, {Kryszczynska, A.}, {Lamy, P.}, {Le Coroller, H.}, {Marciniak, A.},
  {Michalowski, T.}, {Michel, P.}, {Mugnier, L. M.}, {Neichel, B.}, {Pajuelo,
  M.}, {Podlewska-Gaca, E.}, {Santana-Ros, T.}, {Tanga, P.}, {Vachier, F.},
  {Vigan, A.}, {Witasse, O.}, \& {Yang, B.}}]{Fetick-2019a}
{F\'etick, R. JL.}, {Jorda, L.}, {Vernazza, P.}, {et~al.} 2019, A\&A, 623, A6

\bibitem[{Fusco {et~al.}(2004{\natexlab{a}})Fusco, Ageorges, Rousset, Rabaud,
  Gendron, Mouillet, Lacombe, Zins, Charton, Lidman, \& Hubin}]{Fusco-2004b}
Fusco, T., Ageorges, N., Rousset, G., {et~al.} 2004{\natexlab{a}}, in
  Advancements in Adaptive Optics, ed. D.~B. Calia, B.~L. Ellerbroek, \&
  R.~Ragazzoni, Vol. 5490, International Society for Optics and Photonics
  (SPIE), 118 -- 129

\bibitem[{Fusco {et~al.}(2004{\natexlab{b}})Fusco, Rousset, Rabaud, Gendron,
  Mouillet, Lacombe, Zins, Madec, Lagrange, Charton, Rouan, Hubin, \&
  Ageorges}]{Fusco-2004a}
Fusco, T., Rousset, G., Rabaud, D., {et~al.} 2004{\natexlab{b}}, Journal of
  Optics A: Pure and Applied Optics, 6, 585

\bibitem[{{Gendron, E.} {et~al.}(2006){Gendron, E.}, {Cl\'enet, Y.}, {Fusco,
  T.}, \& {Rousset, G.}}]{Gendron-2006}
{Gendron, E.}, {Cl\'enet, Y.}, {Fusco, T.}, \& {Rousset, G.} 2006, A\&A, 457,
  359

\bibitem[{Gilles {et~al.}(2012)Gilles, Correia, V\'{e}ran, Wang, \&
  Ellerbroek}]{Gilles-2012}
Gilles, L., Correia, C., V\'{e}ran, J.-P., Wang, L., \& Ellerbroek, B. 2012,
  Appl. Opt., 51, 7443

\bibitem[{{Infante-Sainz} {et~al.}(2019){Infante-Sainz}, {Trujillo}, \&
  {Rom{\'a}n}}]{Infante2019}
{Infante-Sainz}, R., {Trujillo}, I., \& {Rom{\'a}n}, J. 2019, \mnras, 2729

\bibitem[{{Kamann}(2018)}]{2018ascl.soft05021K}
{Kamann}, S. 2018, {PampelMuse: Crowded-field 3D spectroscopy}

\bibitem[{Kamann {et~al.}(2018{\natexlab{a}})Kamann, Bastian, Husser,
  Martocchia, Usher, den Brok, Dreizler, Kelz, Krajnović, Richard, Steinmetz,
  \& Weilbacher}]{Kamann-2018}
Kamann, S., Bastian, N., Husser, T.-O., {et~al.} 2018{\natexlab{a}}, Monthly
  Notices of the Royal Astronomical Society, 480, 1689

\bibitem[{Kamann {et~al.}(2018{\natexlab{b}})Kamann, Husser, Dreizler,
  Emsellem, Weilbacher, Martens, Bacon, den Brok, Giesers, Krajnović, Roth,
  Wendt, \& Wisotzki}]{Kamann-2017}
Kamann, S., Husser, T.-O., Dreizler, S., {et~al.} 2018{\natexlab{b}}, Monthly
  Notices of the Royal Astronomical Society, 473, 5591

\bibitem[{{Kamann} {et~al.}(2013){Kamann}, {Wisotzki}, \&
  {Roth}}]{2013A&A...549A..71K}
{Kamann}, S., {Wisotzki}, L., \& {Roth}, M.~M. 2013, \aap, 549, A71

\bibitem[{Kolb {et~al.}(2017)Kolb, Madec, Arsenault, Oberti, Paufique, Penna,
  Ströbele, Donaldson, Soenke, Valles, Kiekebusch, Argomedo, Louarn, Vernet,
  Haguenauer, Duhoux, Aller-Carpentier, Valenzuela, \& Guerra}]{Kolb-2017}
Kolb, J., Madec, P.-Y., Arsenault, R., {et~al.} 2017

\bibitem[{Madec {et~al.}(2018)Madec, Arsenault, Kuntschner, Kolb, Pirard,
  Paufique, Penna, Hackenberg, Vernet, Valles, \& Hubin}]{MUSE-WFM-spec}
Madec, P.-Y., Arsenault, R., Kuntschner, H., {et~al.} 2018, in Adaptive Optics
  Systems VI, ed. L.~M. Close, L.~Schreiber, \& D.~Schmidt, Vol. 10703,
  International Society for Optics and Photonics (SPIE), 1 -- 13

\bibitem[{{Moffat}(1969)}]{Moffat1969}
{Moffat}, A.~F.~J. 1969, \aap, 3, 455

\bibitem[{{M{\"u}ller S{\'a}nchez} {et~al.}(2006){M{\"u}ller S{\'a}nchez},
  {Davies}, {Eisenhauer}, {Tacconi}, {Genzel}, \& {Sternberg}}]{Muller2006}
{M{\"u}ller S{\'a}nchez}, F., {Davies}, R.~I., {Eisenhauer}, F., {et~al.} 2006,
  \aap, 454, 481

\bibitem[{Neichel {et~al.}(2009)Neichel, Fusco, \& Conan}]{Neichel2008}
Neichel, B., Fusco, T., \& Conan, J.-M. 2009, J. Opt. Soc. Am. A, 26, 219

\bibitem[{{Oberti} {et~al.}(2018){Oberti}, {Kolb}, {Madec}, {Haguenauer}, {Le
  Louarn}, {Pettazzi}, {Guesalaga}, {Donaldson}, {Soenke}, {Jeram}, {Su{\'a}rez
  Valles}, {Kiekebusch}, {Argomedo}, {La Penna}, {Paufique}, {Arsenault},
  {Hubin}, \& {Vernet}}]{Oberti2018}
{Oberti}, S., {Kolb}, J., {Madec}, P.-Y., {et~al.} 2018, in Society of
  Photo-Optical Instrumentation Engineers (SPIE) Conference Series, Vol. 10703,
  \procspie, 107031G

\bibitem[{{Pirard} {et~al.}(2004){Pirard}, {Kissler-Patig}, {Moorwood},
  {Biereichel}, {Delabre}, {Dorn}, {Finger}, {Gojak}, {Huster}, {Jung}, {Koch},
  {Le Louarn}, {Lizon}, {Mehrgan}, {Pozna}, {Silber}, {Sokar}, \&
  {Stegmeier}}]{Pirard2004}
{Pirard}, J.-F., {Kissler-Patig}, M., {Moorwood}, A., {et~al.} 2004, Society of
  Photo-Optical Instrumentation Engineers (SPIE) Conference Series, Vol. 5492,
  {HAWK-I: A new wide-field 1- to 2.5-{\ensuremath{\mu}}m imager for the VLT},
  ed. A.~F.~M. {Moorwood} \& M.~{Iye}, 1763--1772

\bibitem[{Ragland {et~al.}(2016)Ragland, Jolissaint, Wizinowich, van Dam,
  Mugnier, Bouxin, Chock, Kwok, Mader, Witzel, Do, Fitzgerald, Ghez, Lu,
  Martinez, Morris, \& Sitarski}]{Ragland-2016}
Ragland, S., Jolissaint, L., Wizinowich, P., {et~al.} 2016, in Adaptive Optics
  Systems V, ed. E.~Marchetti, L.~M. Close, \& J.-P. Véran, Vol. 9909,
  International Society for Optics and Photonics (SPIE), 573 -- 590

\bibitem[{{Trujillo} {et~al.}(2001){Trujillo}, {Aguerri}, {Cepa}, \&
  {Guti{\'e}rrez}}]{Trujillo2001}
{Trujillo}, I., {Aguerri}, J.~A.~L., {Cepa}, J., \& {Guti{\'e}rrez}, C.~M.
  2001, \mnras, 328, 977

\bibitem[{V\'{e}ran {et~al.}(1997)V\'{e}ran, Rigaut, Ma\^{i}tre, \&
  Rouan}]{Veran-1997}
V\'{e}ran, J.-P., Rigaut, F., Ma\^{i}tre, H., \& Rouan, D. 1997, J. Opt. Soc.
  Am. A, 14, 3057

\bibitem[{{Weilbacher} {et~al.}(2014){Weilbacher}, {Streicher}, {Urrutia},
  {P{\'e}contal-Rousset}, {Jarno}, \& {Bacon}}]{2014ASPC..485..451W}
{Weilbacher}, P.~M., {Streicher}, O., {Urrutia}, T., {et~al.} 2014,
  Astronomical Society of the Pacific Conference Series, Vol. 485, {The MUSE
  Data Reduction Pipeline: Status after Preliminary Acceptance Europe}, ed.
  N.~{Manset} \& P.~{Forshay}, 451

\bibitem[{Wright {et~al.}(2009)Wright, Larkin, Law, Steidel, Shapley, \&
  Erb}]{wright-2009}
Wright, S.~A., Larkin, J.~E., Law, D.~R., {et~al.} 2009, The Astrophysical
  Journal, 699, 421

\end{thebibliography}
\end{document}